\documentclass[11pt,a4paper]{article}
\usepackage{parskip}
\usepackage[utf8]{inputenc}
\usepackage[T1]{fontenc}
\usepackage{lmodern}
\usepackage[english]{babel}
\usepackage{amsmath,amssymb,amsfonts}
\usepackage{mathtools}
\usepackage{bm}
\usepackage{graphicx}
\usepackage{authblk}
\usepackage{hyperref}
\usepackage[a4paper,margin=2.5cm]{geometry}
\usepackage{enumitem}
\usepackage{setspace}
\usepackage{titlesec}
\usepackage{xcolor}
\usepackage{caption}

\hypersetup{
    colorlinks=true,
    linkcolor=blue,
    citecolor=blue,
    urlcolor=blue
}

\titleformat{\section}{\normalfont\large\bfseries}{\thesection.}{0.5em}{}
\titleformat{\subsection}{\normalfont\normalsize\bfseries}{}{0em}{}

\begin{document}

\title{\textbf{Noise in analog programmable-photonic computation}}

\author[a]{Raúl López-March}
\author[a,*]{Andrés Macho-Ortiz}
\author[b]{Francisco Javier Fraile-Peláez}
\author[a,c,*]{ \protect\\José Capmany}

\affil[a]{ITEAM Research Institute, Universitat Politècnica de València, Valencia, 46022, Spain}
\affil[b]{Dept. Teoría de la Señal y Comunicaciones, Universidad de Vigo E.I. Telecomunicación, Campus Universitario, E-36202 Vigo (Pontevedra), Spain}
\affil[c]{iPronics, Programmable Photonics, S.L, Camino de Vera s/n, Valencia, 46022, Spain}
\affil[*]{Corresponding authors: \href{mailto:amachor@iteam.upv.es}{amachor@iteam.upv.es}, \href{mailto:jcapmany@iteam.upv.es}{jcapmany@iteam.upv.es}}

\date{}

\maketitle

% --- Abstract ---
\noindent\textbf{Abstract}

\noindent Analog Programmable-Photonic Computation (APC) leverages programmable integrated photonics (PIP) to perform high-speed matrix operations using optical waves. However, the continuous nature of optical waves that implement the analog bits or anbits --- the fundamental unit of information in APC --- makes computational results intrinsically sensitive to physical noise. Here, we establish and experimentally validate a comprehensive noise analysis in APC platforms using the geometric representation of the anbit in the Generalized Bloch Sphere (GBS). By modeling the physical noise sources in PIP circuits as random photocurrent fluctuations at the output of the opto-electrical (O/E) converter, and using error propagation theory, the noise statistics can be projected onto the GBS. This approach leads to specific noise maps in the GBS for each noise source, enabling the identification of the dominant noise sources within the APC system. Analytical predictions are numerically and experimentally validated on a fabricated silicon PIP chip. Beyond statistical characterization of system noise, the proposed model provides quantitative design criteria for noise-adapted analog constellations in the GBS, advancing APC towards scalable and robust optical computing systems, with potential applications in emerging paradigms such as photonic neuromorphic computing.

\vspace{0.5em}
\noindent\textbf{Keywords:} programmable integrated photonics, optical computing, noise

\vspace{1em}

% --- 1. Introduction ---
\section{Introduction}

The dominance of digital electronics in computing applications is being challenged by the physical constraints in the miniaturization and power efficiency of electronic transistors marked by Moore's and Dennard's laws [1-3]. These limitations hinder the efficient implementation of emerging computational tasks, especially those that require real-time matrix processing with low latency, high bandwidth and parallelism [4,5]. In this scenario, some representative applications are medical diagnostic imaging [6], artificial intelligence [7], robotics [8], remote sensing [9], lidar [10], and quantum optimization [11], among others.

Consequently, alternative computing models implemented with non-electronic technologies are the scope of intensive research in recent years. In particular, non-digital approaches based on matrix algebra, such as neuromorphic and quantum computing, are explored using technologies with complementary hardware requirements to those of electronics in reconfigurability, scalability, bandwidth, energy consumption, and parallelism [12-15].

Here, programmable integrated photonics (PIP) has emerged as a potential technological candidate, enabling the previous hardware requirements [16-18]. PIP is a system-on-chip platform that exploits the native capability of photonics integrated circuits to implement high-speed matrix signal processing through reconfigurable interference of multiple optical signals. This reconfigurability is achieved by connecting 2$\times$2 circuits (built from phase shifters, beam splitters, beam combiners, and resonators) in an optical mesh [16]. Furthermore, the CMOS compatibility of silicon PIP circuitry enables the coexistence with digital electronics, thus providing an efficient hardware-acceleration solution for computational problems requiring matrix operations [18-20].

As a result, the computational potential of PIP technology has so far been explored primarily through the acceleration of matrix operations within the context of quantum and neuromorphic paradigms [12,15-21]. Nevertheless, the PIP implementation of these computing models has encountered fundamental limitations, as both quantum and neuromorphic computing theories were not originally conceived to exploit the intrinsic capabilities of integrated photonics. Optical quantum computing is extremely sensitive to environmental noise and optical losses of PIP circuits, thus requiring a large number of ancillary qubits to perform error correction and, therefore, limiting scalability [15]. Photonic neuromorphic computing faces challenges mimicking neuron-like activation functions, defined via non-linear operations, which are difficult to implement and scale efficiently using PIP circuitry [22,23].

\begin{figure}[h!]
    \centering
    \includegraphics[width=0.9\textwidth]{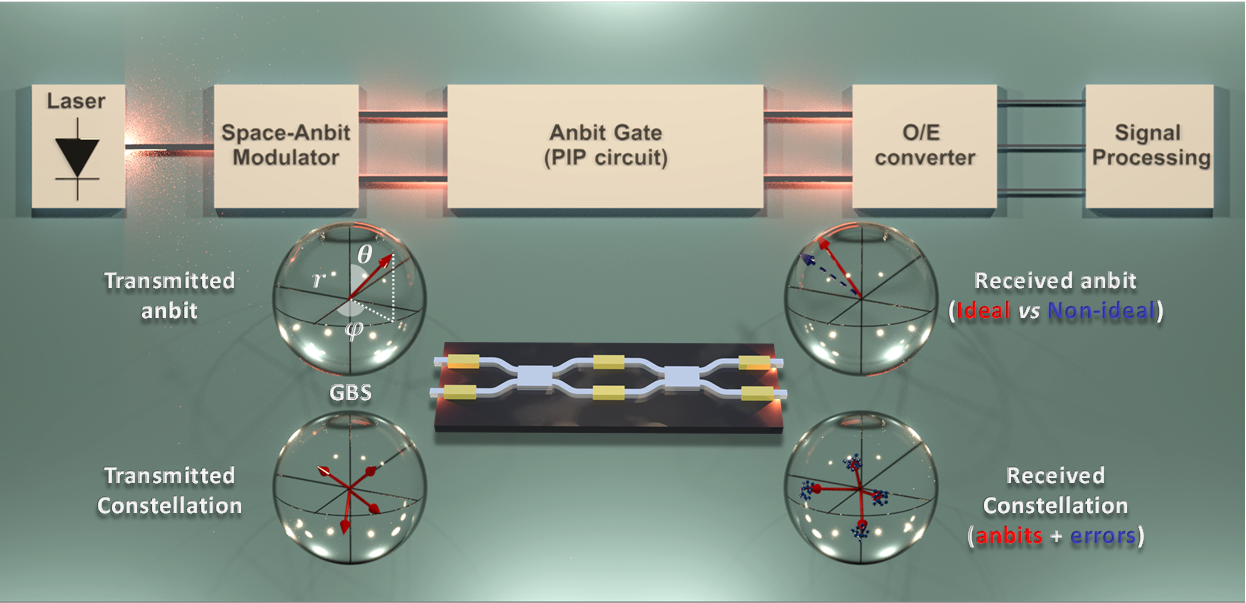}
    \caption{\textbf{Conceptual structure of an analog programmable-photonic computing (APC) system [24,25].} Any computational system is an information-processing system encompassing a transmitter, a channel that performs computational operations, and a receiver. From this perspective, an APC system is composed of (\emph{i}) the laser and the modulator that generate the unit of information -- the analog bit or anbit, whose state is geometrically represented as a point on the Generalized Bloch Sphere (GBS) with spherical coordinates $r,\theta,\varphi$, also termed as the effective degrees of freedom (EDF) of the anbit; (\emph{ii}) programmable integrated photonic (PIP) circuits implementing anbit gates, which compute the state of the anbit by modifying its location in the GBS (e.g. performing rotations via unitary matrices); and (\emph{iii}) the opto-electrical (O/E) converter that maps the output states -- the ideal states (red arrows) along with errors (blue points) -- from the optical to the electrical domain.}
    \label{fig:fig1}
\end{figure}

In this context, a new computation theory -- referred to as Analog Programmable-Photonic Computation (APC) -- has been introduced in ref. [24]. Notably, APC model is specifically adapted to PIP hardware, leveraging the classical analog nature of optical waves along with the ability of PIP to perform matrix-by-vector multiplications. Accordingly, its unit of information -- the analog bit or anbit, a two-dimensional vector -- and its basic operations have been defined with computational properties aligned with the technological features of PIP. Interestingly, the simplicity of this new paradigm relies on the fact that the states of the anbit, its operations, and errors (arising from system noise and hardware imperfections) can be intuitively visualized using a geometrical representation -- the \emph{Generalized Bloch Sphere} (GBS), see Fig.~\ref{fig:fig1}.

Furthermore, its accompanying information theory -- termed Analog Programmable-Photonic Information (API) -- has been recently presented in ref. [25] to complement APC with a mathematical framework that describes how user information should be generated, computed and recovered to minimize the computational errors. To this end, API proposes the mitigation of such errors by engineering the generated information via a discrete constellation of points in the GBS, which define the set of allowed anbit states in the computational system (Fig.~\ref{fig:fig1}). The optimization of this ``analog constellation'' at the transmitter allows us to minimize the error probability at the receiver, making APC robust against computational errors. As a result, error correction becomes a non-critical component in this computational landscape.

Nevertheless, the practical realization of any analog computing approach such as APC introduces fundamental challenges compared to digital systems. Unlike digital signals, which offer intrinsic robustness against small perturbations due to their discrete nature [26], analog waves are continuously valued and thus more sensitive to physical impairments such as noise and hardware imperfections [17,26]. Hence, in APC, these impairments can be accumulated when the anbit states are propagated through the PIP circuits, degrading the optical performance of the computational system. Although this issue -- intrinsic to any analog computing model -- is circumvented in APC via the discretization of the GBS [25], thereby defining the corresponding analog constellation and endowing the states with a discrete nature (akin to the symbols of a digital constellation [27]), \emph{it remains crucial to analyze how system noise is distributed in the GBS}, as this will guide the design of optimal analog constellations.

In this work, we establish a comprehensive and experimentally validated model for noise analysis in APC. The paper is organized as follows. First, we present the theoretical framework required to analyze noise statistics in APC systems implemented with PIP circuitry. Next, we identify and characterize the dominant noise sources by combining our theoretical model with numerical simulations and experimental measurements. Beyond characterizing noise, the framework provides quantitative criteria to design noise-adapted analog constellations in the GBS. Finally, we outline future research directions toward scalable and robust APC systems, including their potential application to emerging paradigms such as photonic neuromorphic computing.

% --- 2. Theory ---
\section{Theory}

Following the principles of API theory [25], any APC platform can be regarded as an information-processing system composed of a transmitter that generates the information, a channel that performs the required computational operations using PIP circuitry, and a receiver that recovers the computational results (see Fig.~\ref{fig:fig1}). Each of these three blocks contributes distinct noise sources that are, or can be accurately approximated as, zero-mean wide-sense stationary Gaussian processes. At the transmitter, the laser introduces \emph{relative intensity noise} (RIN) and \emph{phase noise} [28,29], while the space-anbit modulator, implemented with a reconfigurable Mach-Zehnder interferometer [24,25], adds additional \emph{phase noise} through its integrated phase shifters [18,30]. At the channel, additional \emph{phase noise} originates from the phase shifters in the PIP circuits that implement the anbit gates. As detailed in [24], these computational operations can be realized without optical amplification stages, thereby avoiding undesired nonlinear and lasing effects in feedback-based PIP architectures. Accordingly, the absence of optical amplifiers within the channel precludes the presence of amplified spontaneous emission (ASE) noise [31]. Finally, at the receiver, \emph{shot noise} -- originating from the discrete nature of photodetection events [29] -- and \emph{thermal noise} -- arising from thermal charge fluctuations in the opto-electrical (O/E) converter [29] -- perturb the information encoded into the anbits.

To assess the impact of the aforementioned noise contributions on the computational platform, we focus our analysis on the final stage of the API system: the O/E converter. This block transforms the received optical field at the channel output into electrical currents, while inevitably transferring not only information encoded into the anbit states, but also the whole contribution of the system noise.

\begin{figure}[h!]
    \centering
    \includegraphics[width=0.9\textwidth]{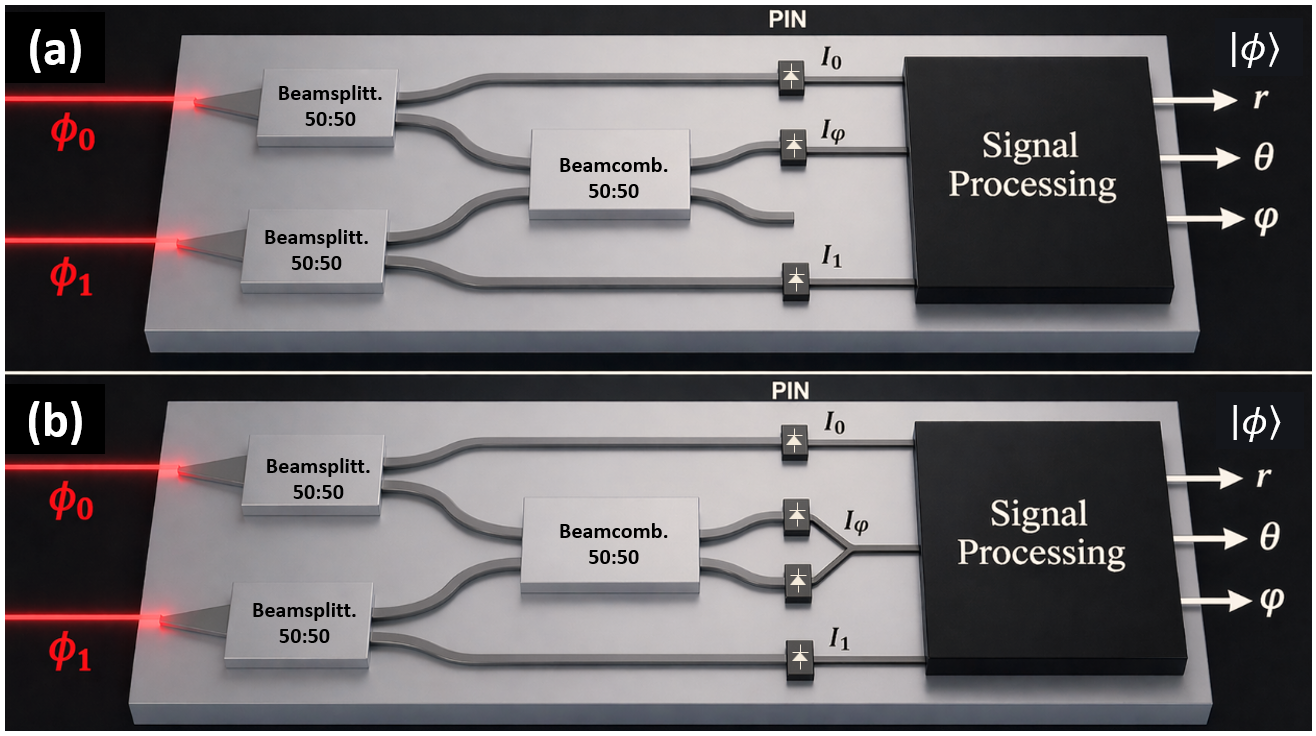}
    \caption{\textbf{Differential O/E converters.} (a) Unbalanced architecture. (b) Balanced architecture. The 50:50 beam splitters are implemented with Y-junctions, while the 50:50 beam combiners are realized using multi-mode interferometers.}
    \label{fig:fig2}
\end{figure}

Along this line, in order to elucidate the basic behavior of the system noise and its distribution within the GBS at the output of the O/E converter, we start our analysis by considering an API system in its most elementary configuration, depicted in Fig.~\ref{fig:fig1}: a single transmitter emitting a sequence of states, a channel composed of a 2$\times$2 unitary PIP circuit that computes each state independently through a single-anbit unitary gate (U-gate) [24,32], and a receiver using \emph{unbalanced} differential O/E conversion, whose hardware is shown in Fig.~\ref{fig:fig2}(a). This API system is of special interest in APC due to its minimal hardware implementation and energy efficiency [25]. Additional intricate scenarios with feedback architectures, multi-anbit gates, and different classes of O/E converters will be addressed later in the Discussion section.

In this API system, the optical field at the input of the O/E converter is a two-dimensional analog signal composed of two complex envelopes $\phi_0=|\phi_0|e^{j\arg\phi_0}$ and $\phi_1=|\phi_1|e^{j\arg\phi_1}$ with a phase delay (or differential phase) $\varphi=\arg\phi_1-\arg\phi_0$. Thus, the corresponding state of the anbit is:
\begin{equation}
|\phi\rangle = |\phi_0||0\rangle + e^{j\varphi}|\phi_1||1\rangle \equiv r \left(\cos\frac{\theta}{2}|0\rangle + e^{j\varphi}\sin\frac{\theta}{2}|1\rangle\right),
\label{eq:1}
\end{equation}
where $r$ (radius), $\theta$ (elevation angle), and $\varphi$ (azimuthal angle) are referred to as the \emph{effective degrees of freedom} (EDFs), which define the \emph{spherical coordinates} of a position vector in the GBS, denoted $\mathbf{r}\equiv(r,\theta,\varphi)$ [24,25]. The unbalanced differential O/E converter recovers the moduli $|\phi_0|$ and $|\phi_1|$ from the complex envelopes $\phi_0$ and $\phi_1$ of the optical field via direct detection and extracts the differential phase $\varphi$ inducing an interference between $\phi_0$ and $\phi_1$ using a 50:50 beam combiner. The resulting photocurrents $I_0$, $I_1$, and $I_\varphi$ are given by the following expressions [25]:
\begin{equation}
I_0 = \tfrac{1}{2}\mathcal{R}|\phi_0|^2, \quad I_1 = \tfrac{1}{2}\mathcal{R}|\phi_1|^2, \quad I_\varphi = \tfrac{1}{4}\mathcal{R}\left(|\phi_0|^2 + |\phi_1|^2 - 2|\phi_0||\phi_1|\sin\varphi\right),
\label{eq:2}
\end{equation}
with $\mathcal{R}$ being the responsivity of the photodiode. Hence, the EDFs defining the received state $|\phi\rangle$ are:
\begin{equation}
r = \sqrt{\frac{2(I_0+I_1)}{\mathcal{R}}}, \quad \theta = 2\arctan\sqrt{\frac{I_1}{I_0}}, \quad \varphi = \arcsin\left(\frac{I_0+I_1-2I_\varphi}{2\sqrt{I_0 I_1}}\right).
\label{eq:3}
\end{equation}

Under ideal conditions, Eq.~(\ref{eq:3}) defines a single point (or state) in the GBS via the deterministic position vector $\mathbf{r}$. However, in practice, the distinct noise sources of the system induce random temporal fluctuations in the photocurrents, leading to a position vector with a \emph{random} behavior. Specifically, in a \emph{noisy} API system, $I_0(t)$, $I_1(t)$, and $I_\varphi(t)$ are random processes, which may be assumed stationary, consistent with the stationary nature of the noise sources mentioned at the beginning of this section. Therefore, the (time-independent) probabilistic distributions of the photocurrents are captured by the probability density functions (pdf) of the EDFs $r$, $\theta$, and $\varphi$; which can be regarded as random variables instead of random processes in the GBS (a time-independent geometrical picture in APC [24]).

In this way, the ideal state that should be detected in noiseless conditions, along with the overall noise contributions, is visualized in the GBS as a \emph{three-dimensional (3D) region}, defining the range of the random vector $\mathbf{r}$ (Fig.~\ref{fig:fig1}). As the system noise increases, the volume of the 3D region in the GBS encompassing the ideal state and its noise contributions expands. As a result, in an analog constellation defined with multiple ideal states at the transmitter, the distance between neighboring 3D noisy regions in the received constellation decreases, increasing the error probability at the receiver [25].

This degradation of the system performance may have a pronounced impact on the differential phase $\varphi$, as $I_\varphi$ inherits noise from the common-mode terms $|\phi_0|^2$ and $|\phi_1|^2$. To mitigate this effect, we propose a \emph{balanced} differential O/E conversion architecture, see Fig.~\ref{fig:fig2}(b). This configuration resembles the unbalanced scheme, but includes a second PIN photodiode connected to the previously unused output port of the 2$\times$2 combiner. In this way, the photocurrents $I_0$ and $I_1$ are governed by the same expressions as those of the unbalanced converter, thus preserving the statistics of the EDFs $r$ and $\theta$. Nonetheless, the photocurrent $I_\varphi$ and the differential phase $\varphi$ are now found to be (Supplementary Section 1):
\begin{equation}
I_\varphi = -\mathcal{R}|\phi_0||\phi_1|\sin\varphi,
\label{eq:4}
\end{equation}
\begin{equation}
\varphi = \arcsin\left(\frac{-I_\varphi}{2\sqrt{I_0 I_1}}\right).
\label{eq:5}
\end{equation}

Comparing $I_\varphi$ in Eqs.~(\ref{eq:2}) and (\ref{eq:4}), we can infer that the fluctuations (i.e. variance) of $\varphi$ will be reduced in the balanced scheme, as $|\phi_0|^2$ and $|\phi_1|^2$ are removed from $I_\varphi$.

As shown, a statistical characterization of the noise perturbing the EDFs $r$, $\theta$, and $\varphi$ is essential, as it enables us to describe the 3D noisy regions of an analog constellation at the receiver. This characterization requires analyzing the pdf, mean and variance of each EDF. The description of the pdf and mean is especially simple since, as demonstrated in [25], the random fluctuations of the EDFs can be treated as zero-mean Gaussian random variables, consistent with the stochastic nature of the main noise sources present in our computational system. However, \emph{estimating the variance is non-trivial, yet essential for understanding how the 3D noisy regions distribute over the GBS and broaden as the system noise increases}. Therefore, we now derive closed-form expressions for estimating the variance of the EDFs as a function of the statistical parameters of the photocurrents $I_0(t)$, $I_1(t)$, and $I_\varphi(t)$.

To this end, we express each photocurrent $I_k(t)$ ($k\in\{0,1,\varphi\}$) as the sum of its ideal value $\overline{I_k}$, which corresponds to the time-averaged photocurrent (assuming ergodicity), and a zero-mean fluctuation $\delta I_k(t)$ that captures all zero-mean noise contributions, that is, $I_k(t)=\overline{I_k}+\delta I_k(t)$. Next, assuming weak perturbations in $I_k(t)$, in line with the results of [25], we can safely regard the first-order derivative as dominant over higher-order terms. This justifies the use of \emph{first-order error propagation theory} [33] to estimate the variance of the EDFs from the variance of the photocurrents. Using this approach, we find the variances of $r$, $\theta$, and $\varphi$ in the \emph{unbalanced} differential O/E converter (see Supplementary Section 2):
\begin{equation}
\mathrm{Var}(r) \approx \frac{1}{\mathcal{R}^2 \overline{r}^2}\left[\mathrm{Var}(\delta I_0) + \mathrm{Var}(\delta I_1)\right],
\label{eq:6}
\end{equation}
\begin{equation}
\mathrm{Var}(\theta) \approx \frac{4}{\mathcal{R}^2 \overline{r}^4}\left[\tan^2\frac{\overline{\theta}}{2}\mathrm{Var}(\delta I_0) + \cot^2\frac{\overline{\theta}}{2}\mathrm{Var}(\delta I_1)\right],
\label{eq:7}
\end{equation}
\begin{multline}
\mathrm{Var}(\varphi) \approx \frac{4}{\mathcal{R}^2 \overline{r}^4 \sin^2\overline{\theta} \cos^2\overline{\varphi}} \Bigg[\left(1-\sin\overline{\varphi}\tan\frac{\overline{\theta}}{2}\right)^2 \mathrm{Var}(\delta I_0) \\
+ \left(1-\sin\overline{\varphi}\cot\frac{\overline{\theta}}{2}\right)^2 \mathrm{Var}(\delta I_1) + 4\,\mathrm{Var}(\delta I_\varphi)\Bigg].
\label{eq:8}
\end{multline}

Here, $\overline{r}$, $\overline{\theta}$ and $\overline{\varphi}$ denote the ideal values of the EDFs, obtained from the ideal photocurrents $\overline{I_0}$, $\overline{I_1}$, and $\overline{I_\varphi}$ via Eq.~(\ref{eq:3}). In the \emph{balanced} differential O/E converter, the variances of $r$ and $\theta$ remain identical to those obtained in the unbalanced configuration, whereas the variance of $\varphi$ becomes:
\begin{equation}
\mathrm{Var}(\varphi) \approx \frac{1}{\mathcal{R}^2 \overline{r}^4 \cos^2\overline{\varphi}}\left[\left(\frac{\sin\overline{\varphi}}{\cos^2\frac{\overline{\theta}}{2}}\right)^2 \mathrm{Var}(\delta I_0) + \left(\frac{\sin\overline{\varphi}}{\sin^2\frac{\overline{\theta}}{2}}\right)^2 \mathrm{Var}(\delta I_1) + \frac{4}{\sin^2\overline{\theta}}\mathrm{Var}(\delta I_\varphi)\right],
\label{eq:9}
\end{equation}
with $\overline{\varphi}$ now calculated from the ideal photocurrents using Eq.~(\ref{eq:5}). Remarkably, Eqs.~(\ref{eq:6})-(\ref{eq:9}) will allow us to analyze how the overall system noise or any specific noise source distributes over the GBS, as they only depend on: (\emph{i}) the ideal location of the anbit $\overline{\mathbf{r}}\equiv(\overline{r},\overline{\theta},\overline{\varphi})$, and (\emph{ii}) the photocurrent fluctuations $\mathrm{Var}(\delta I_k)$, which can either be experimentally measured (capturing all noise contributions) or estimated for a particular noise source using established models from the literature (e.g., in the case of laser RIN, semi-classical models predict $\mathrm{Var}(\delta I_k^{\mathrm{RIN}}) \propto \overline{I_k}^2$ [28,29]). Therefore, using these expressions, we can examine the theoretical behavior of the variance of each EDF over the GBS under the dominant noise sources in the API system of Fig.~\ref{fig:fig1}: RIN, shot, and thermal noises. Additional noise sources such as the phase noise induced by the laser and the phase shifters are found to be negligible, see Supplementary Section 3.

Figures~\ref{fig:fig3}(a-d) show the variance of each EDF across a GBS at constant radius, illustrating how the combined contributions of RIN, shot noise, and thermal noise distribute across the sphere. These graphics were obtained from Eqs.~(\ref{eq:6})-(\ref{eq:9}) combined with the standard expressions of $\mathrm{Var}(\delta I_k)$ for RIN, thermal and shot noise reported in the literature (see Supplementary Section 4). All variances are normalized to their respective maximum values, enabling comparison independent of system-specific parameters such as the photodiode responsivity. The symmetry of the resulting maps indicates that evaluating the complete GBS is unnecessary. Instead, it is sufficient to consider some \emph{representative regions}, specifically, the plane $\overline{\varphi}=0$ to analyze $\mathrm{Var}(r)$ and $\mathrm{Var}(\theta)$, and the plane $\overline{\theta}=\pi/2$ to analyze $\mathrm{Var}(\varphi)$.

Figures~\ref{fig:fig3}(e-h) illustrate the variance of each EDF over these representative regions, allowing the individual contributions of each noise source to be examined. As RIN, shot, and thermal noise contributions may differ by several orders of magnitude within the API system, all variances are normalized to their respective maximum values. This reveals that each noise source imposes a distinct ``statistical signature'' on the variance of the EDFs.

\begin{figure}[h!]
    \centering
    \includegraphics[width=0.95\textwidth]{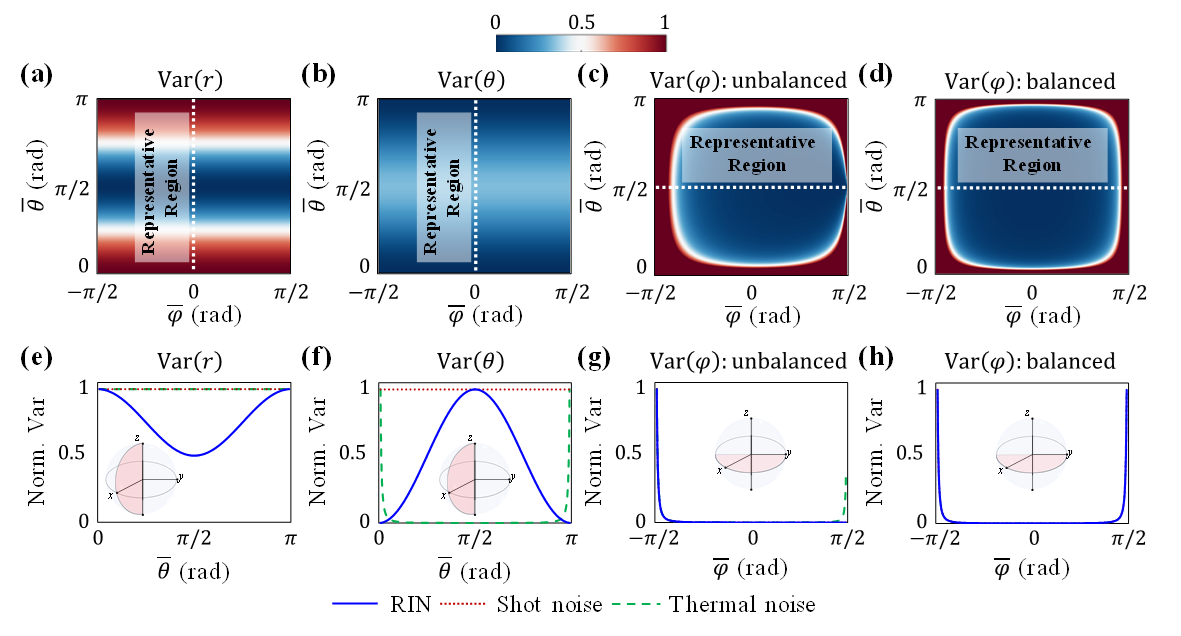}
    \caption{\textbf{Normalized theoretical variance of the effective degrees of freedom (EDF) over a generalized Bloch sphere (GBS) at constant radius under the contributions of RIN, shot, and thermal noise.} (a--d) The colormaps show the normalized variance of the EDFs as a function of the ideal elevation angle ($\overline{\theta}$) and the ideal azimuthal angle ($\overline{\varphi}$) under the \emph{combined} contributions of RIN, shot, and thermal noise. These colormaps are calculated by combining Eqs.~(\ref{eq:6})-(\ref{eq:9}) with the standard expressions of $\mathrm{Var}(\delta I_k)$ for RIN, thermal and shot noise reported in the literature (see Supplementary Section 4 for more details). (e--h) Normalized theoretical variance of the EDFs under the \emph{independent} contribution of RIN, shot, and thermal noise over representative regions of the GBS identified in the colormaps. The representative regions of the GBS that are theoretically analyzed are inserted into each panel (highlighted in red). (e) $\mathrm{Var}(r)$ as a function of $\overline{\theta}$ setting $\overline{\varphi}=0$. (f) $\mathrm{Var}(\theta)$ as a function of $\overline{\theta}$ setting $\overline{\varphi}=0$. (g, h) $\mathrm{Var}(\varphi)$ as a function of $\overline{\varphi}$ taking $\overline{\theta}=\pi/2$, shown for the unbalanced (g) and balanced (h) differential O/E converter.}
    \label{fig:fig3}
\end{figure}

First, the \emph{RIN source} presents a characteristic harmonic profile in $\mathrm{Var}(r)_{\mathrm{RIN}}$ and $\mathrm{Var}(\theta)_{\mathrm{RIN}}$ when varying the elevation angle, which induces respectively a local minimum and a local maximum at $\overline{\theta}=\pi/2$ (the equator of the GBS). In particular, this harmonic behavior can be easily predicted for $\mathrm{Var}(r)_{\mathrm{RIN}}$ by using Eq.~(\ref{eq:6}) noting that: (\emph{i}) $\mathrm{Var}(r)_{\mathrm{RIN}} \propto \mathrm{Var}(\delta I_0)_{\mathrm{RIN}}+\mathrm{Var}(\delta I_1)_{\mathrm{RIN}}$, (\emph{ii}) $\mathrm{Var}(\delta I_k)_{\mathrm{RIN}}\propto \overline{I_k}^2$, and (\emph{iii}) $\overline{I_0}^2+\overline{I_1}^2\propto 1-(1/2)\sin^2\overline{\theta}$ (see Supplementary Section 4 for further details). Moreover, in $\mathrm{Var}(\varphi)_{\mathrm{RIN}}$, the analytical expressions Eqs.~(\ref{eq:8}) and (\ref{eq:9}) predict poles at $\overline{\varphi}=\pm\pi/2$ originating from the term $\cos^2\overline{\varphi}$ in the denominator, defining regions in the GBS where the signal-to-noise ratio is intrinsically degraded. Notably, both singularities are present in the balanced converter, whereas in the unbalanced configuration the pole $\overline{\varphi}=\pi/2$ is not present. Second, \emph{shot noise} presents a flat profile in $\mathrm{Var}(r)_{\mathrm{shot}}$ and $\mathrm{Var}(\theta)^{\mathrm{shot}}$. Since shot noise is linearly proportional to the photocurrent [29], then $\mathrm{Var}(\delta I_k)^{\mathrm{shot}}\propto \overline{I_k}$, which inserted into Eqs.~(\ref{eq:6}) and (\ref{eq:7}) lead to a constant variance in $\mathrm{Var}(r)_{\mathrm{shot}}$ and $\mathrm{Var}(\theta)_{\mathrm{shot}}$ with respect to the elevation angle. Furthermore, shot noise exhibits the same profile as the RIN when analyzing $\mathrm{Var}(\varphi)_{\mathrm{shot}}$ for both unbalanced and balanced converters. Finally, in contrast to RIN and shot noises, \emph{thermal noise} is independent of $\overline{I_k}$ [29]. This implies that the behavior of $\mathrm{Var}(r)_{\mathrm{thermal}}$, $\mathrm{Var}(\theta)_{\mathrm{thermal}}$, and $\mathrm{Var}(\varphi)_{\mathrm{thermal}}$ through the GBS is described by Eqs.~(\ref{eq:6})-(\ref{eq:9}) by taking constant the terms $\mathrm{Var}(\delta I_k)$. As a result, thermal noise is expected to increase at the GBS positions where at least one of the photocurrents heavily decreases, concretely, in the poles. As demonstrated in the next section, these ``statistical signatures'' provide a theoretical tool for identifying the dominant noise source by experimentally analyzing the spatial profiles of $\mathrm{Var}(r)$, $\mathrm{Var}(\theta)$, and $\mathrm{Var}(\varphi)$ over the representative regions of the GBS shown in Fig.~\ref{fig:fig3}.

% --- 3. Results ---
\section{Results}

\begin{figure}[h!]
    \centering
    \includegraphics[width=0.95\textwidth]{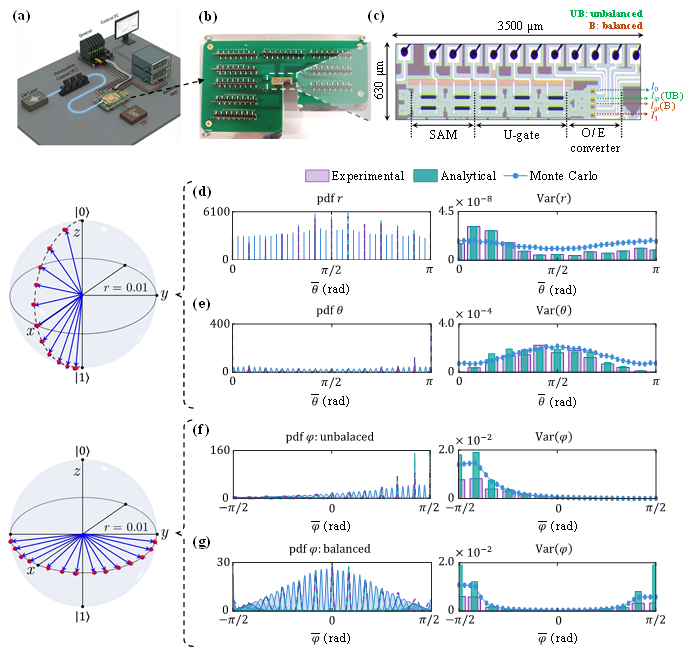}
    \caption{\textbf{Experimental results.} (a) Scheme of the laboratory set-up. A continuous-wave (CW) laser operating at 1550~nm is coupled into the programmable integrated photonic (PIP) circuit through vertical grating couplers after polarization control (PC). Photocurrents at the output of the PIP circuit are measured using a high-precision source-measure unit (SMU). (b) Packaged chip, including the electrical printed circuit board, wiring, and photonic chip. (c) Micrograph of the fabricated PIP chip, which integrates a space-anbit modulator (SAM) [24], a universal unitary gate (U-gate) [24,32], and a differential O/E converter operating in both unbalanced (UB) and balanced (B) configurations. (d--g) Experimental statistics of the effective degrees of freedom (EDF) over the generalized Bloch sphere (GBS) under the combined contributions of RIN, shot, and thermal noise. (d) Probability density function (pdf) and variance of the radius ($r$) as a function of the ideal elevation angle ($\overline{\theta}$) setting the ideal azimuthal angle ($\overline{\varphi}$) to the value $\overline{\varphi}=0$. (e) Pdf and variance of $\theta$ as a function of $\overline{\theta}$ setting $\overline{\varphi}=0$. (f,g) Pdf and variance of $\varphi$ as a function of $\overline{\varphi}$ taking $\overline{\theta}=\pi/2$, shown for the unbalanced (f) and balanced (g) differential O/E converter. The representative regions of the GBS that are experimentally analyzed are inserted at the left of each panel.}
    \label{fig:fig4}
\end{figure}

\begin{figure}[h!]
    \centering
    \includegraphics[width=0.95\textwidth]{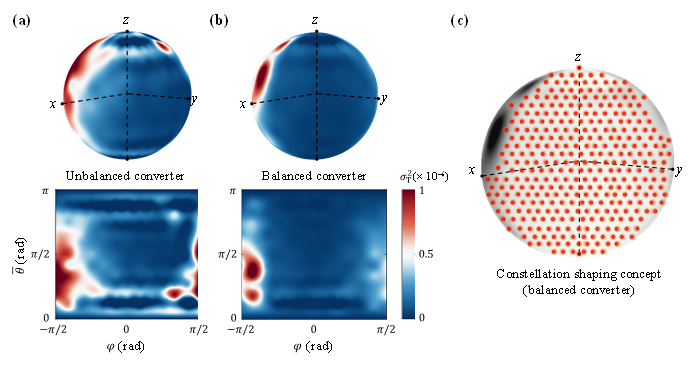}
    \caption{\textbf{Experimental distribution of the total angular variance ($\sigma_{\mathrm{T}}^2) $ over the generalized Bloch sphere (GBS).} Multiple states were generated and recovered on a constant-radius GBS using both unbalanced and balanced differential O/E converters. The total angular variance associated with each recovered state was measured and mapped onto the colormaps shown in (a) for the unbalanced configuration and in (b) for the balanced configuration. (c) Conceptual illustration of noise-adapted constellation shaping on the GBS for the balanced case. Inspired by probabilistic constellation shaping techniques [34], the spatial location of the states (red points) is adapted according to their probability of occurrence, such that higher-probability states are placed in regions of lower noise.}
    \label{fig:fig5}
\end{figure}

In this section, we present an experimental and numerical validation of the theoretical framework developed in Section 2. We characterize the statistical behavior of the EDFs over the GBS under realistic physical conditions and validate the APC/API noise model by comparing the analytical expressions provided by Eqs.~(\ref{eq:6})-(\ref{eq:9}) with experimental measurements and Monte Carlo simulations. Figures~\ref{fig:fig4} and \ref{fig:fig5} summarize the key results of the experimental and numerical work.

\noindent\textbf{Experimental set-up.} The experimental measurements are performed using the laboratory set-up depicted in Fig.~\ref{fig:fig4}(a), which implements the API system sketched in Fig.~\ref{fig:fig1}. A tunable continuous-wave laser (TUNICS T100R), operating at 1550~nm with a linewidth of 400~kHz and a nominal RIN of $-145$~dB/Hz, is coupled to the PIP circuit shown in Fig.~\ref{fig:fig4}(b,c). The circuit integrates a space-anbit modulator (a reconfigurable Mach-Zehnder interferometer [24,25]), a universal single-anbit U-gate [24,32], and a differential O/E converter operating in both unbalanced and balanced configurations. The photocurrents generated by the O/E converter are measured using a precision source-measure unit (Keithley 2400 SourceMeter). The PIP circuit was fabricated by Advanced Micro Foundry using a silicon-on-insulator platform. Additional details regarding chip fabrication and characterization are provided in Supplementary Section 5.

\noindent\textbf{Methodology.} Multiple anbit states $|\phi\rangle = |\phi_0||0\rangle + e^{j\varphi}|\phi_1||1\rangle$ are generated at the transmitter using continuous-wave complex envelopes $\phi_0$ and $\phi_1$ with a phase delay $\varphi=\arg(\phi_1)-\arg(\phi_0)$. These states are located in the same representative regions of the GBS analyzed in Fig.~\ref{fig:fig3} to enable a direct comparison between theoretical and experimental results. The generated states $|\phi\rangle$ propagate through the U-gate (programmed as the identity matrix to avoid unnecessary rotations of the states in the GBS) and are subsequently recovered at the receiver by measuring 300 temporal samples in each photocurrent $I_0(t)$, $I_1(t)$, and $I_\varphi(t)$ at the output of the O/E converter. The photocurrent $I_\varphi$ associated to the unbalanced and balanced configuration is detailed in Fig.~\ref{fig:fig4}(c). Moreover, to ensure uncorrelation between consecutive samples in each photocurrent, the sampling period ($\sim$ms) was chosen to exceed the correlation time of the laser RIN ($\sim 0.01$~$\mu$s). The acquired samples are mapped onto the GBS using Eqs.~(\ref{eq:3}) and (\ref{eq:5}), from which the statistical parameters of the three EDFs $r$, $\theta$, and $\varphi$ are extracted for each generated state (Figs.~\ref{fig:fig4}(d-g)).

For completeness, the experimental sampling procedure is reproduced numerically via a Monte Carlo simulation implemented in MATLAB software. The physical models used to incorporate RIN, shot, and thermal noise into the simulated photocurrents are identical to those employed in Fig.~\ref{fig:fig3}, also detailed in Supplementary Section 4.

\noindent\textbf{Experimental and numerical data.} The received states are characterized by EDFs whose stochastic fluctuations can be accurately modeled as zero-mean Gaussian random variables. The zero-mean assumption is experimentally validated by verifying that the measured mean values of the EDFs coincide with the corresponding ideal values that are expected to be received. Moreover, the statistical signatures shown in Figs.~\ref{fig:fig4}(d-g) demonstrate a strong agreement with the normalized theoretical predictions shown in Fig.~\ref{fig:fig3}. This consistency indicates that \emph{RIN is the dominant noise source in the experimental platform}, exceeding both shot and thermal noise under the operating conditions considered.

A closer look at Figs.~\ref{fig:fig4}(d-g) allows us to extract several quantitative observations. Firstly, the radius ($r$) exhibits a variance on the order of $\sim 10^{-8}$, which is significantly lower than that of the angular EDFs. This confirms that the radial coordinate is intrinsically robust against system noise, reflecting the stability of unitary operations that preserve optical power over the sphere. Secondly, the variance of the elevation angle ($\theta$) reaches values on the order of $\sim 10^{-4}$ at the equator and decreases toward the poles, reproducing the characteristic RIN-induced profile predicted in Fig.~\ref{fig:fig3}(f). This spatial dependence directly follows from the harmonic behavior imposed by the RIN contribution, as predicted by our analytical model in Section 2. Thirdly, the azimuthal angle ($\varphi$) is identified as the most noise-sensitive EDF. Its variance reaches a maximum on the order of $\sim 10^{-3}$, occurring around $\overline{\varphi}\approx -\pi/2$ in the unbalanced converter and around $\overline{\varphi}\approx \pm\pi/2$ in the balanced converter, in agreement with the theoretical trends shown in Fig.~\ref{fig:fig3}(g,h).

Along this line, note that the experimental measurements and Monte Carlo simulations closely match the analytical predictions of the proposed noise model. The slight deviations observed between experimental data and both the simulations and the closed-form expressions in Eqs.~(\ref{eq:6})-(\ref{eq:9}) are likely influenced by residual optical power fluctuations, potentially arising from fiber-to-chip coupling instabilities, which are not explicitly included in the simulation parameters nor captured by the analytical framework.

Finally, it is worth mentioning that the main difference between the unbalanced and balanced converters becomes evident when comparing the \emph{total angular variance} $\sigma_{\mathrm{T}}^2$ (expressed in rad$^2$), induced by all noise contributions across the GBS, see Figs.~\ref{fig:fig5}(a,b). This total angular variance quantifies the accumulated noise-induced errors in both the elevation and azimuthal angles and provides a quantitative performance parameter for assessing the relative robustness of unbalanced and balanced O/E conversion architectures. Moreover, it admits a clear geometric interpretation. Specifically, when the radius is fixed, noise-induced perturbations are approximately confined to the surface of the GBS since $\mathrm{Var}(r)\sim 10^{-8}$. In this case, the total angular variance can be geometrically interpreted as the \emph{metric line element} on a sphere of constant radius ($\overline{r}\approx 0.01$), leading to the expression $\sigma_{\mathrm{T}}^2= \overline{r}^2\mathrm{Var}(\theta) + \overline{r}^2\sin^2\overline{\theta}\,\mathrm{Var}(\varphi)$ (Supplementary Section 6). As shown in Fig.~\ref{fig:fig5}(b), the balanced differential O/E converter yields a significantly more uniform distribution of the total angular variance over the GBS. This improved uniformity in the balanced architecture is a direct consequence of the suppression of common-mode noise terms in Eq.~(\ref{eq:4}).

% --- 4. Discussion ---
\section{Discussion}

The results presented in this work establish an experimentally validated model for analyzing noise in APC systems implemented with PIP technology. By modeling the dominant physical noise sources in passive PIP circuits --- RIN, shot, and thermal noise --- as random fluctuations in the photocurrents of the O/E converter, and propagating these fluctuations to the GBS via first-order error-propagation theory, we have derived closed-form expressions for the variances of the EDFs defining the anbit. This approach enables a direct mapping between photocurrent fluctuations and the 3D noisy regions that characterize each ideal state in the received analog constellation of the GBS. The analytical predictions were validated through Monte Carlo simulations and experimental measurements, demonstrating that the stochastic fluctuations of the EDFs can be accurately modeled as zero-mean Gaussian random variables under realistic operating conditions. Beyond characterizing noise, the model provides a practical methodology to compare O/E converter architectures --- here the unbalanced and balanced differential schemes --- and to quantify their performance through geometrically meaningful figures of merit such as the total angular variance. As such, this work lays the statistical foundations required to design robust APC and API systems under realistic physical constraints.

\noindent\textbf{Noise-optimized analog constellation design.} Within the API context, the encoder maps discrete symbols onto a finite set of allowed anbit states, defining an analog constellation in the GBS. The robustness of the computational platform critically depends on how these states are arranged relative to the noise-induced 3D regions at the receiver [25]. The noise maps derived in this work (Figs.~\ref{fig:fig3}, \ref{fig:fig4}, and \ref{fig:fig5}) therefore provide a direct design toolbox for analog constellation engineering.

A first important observation concerns the non-uniform distribution of noise over the GBS. In particular, the angular variances exhibit characteristic spatial signatures under RIN-dominated conditions: harmonic behavior in the elevation angle and singular regions in the azimuthal angle (Fig.~\ref{fig:fig3}). These results suggest that constellation density should not be uniform over the sphere. Regions where the total angular variance is large --- particularly around specific elevation angles depending on the receiver architecture --- should host a lower density of states to minimize state overlap and detection errors. Conversely, regions with intrinsically lower total angular variance can accommodate a higher state density (Fig.~\ref{fig:fig5}).

However, the poles of the GBS deserve special consideration. Although certain angular variances decrease toward the poles, thermal-noise-induced behavior may increase in those regions due to reduced photocurrent levels. Importantly, near the poles the angular coordinates may lose sensitivity, while the radial coordinate remains well defined and intrinsically robust, as demonstrated experimentally by its significantly lower variance compared to the angular EDFs. This opens an interesting design strategy: instead of densely distributing states in angular coordinates near the poles, one can exploit the radial dimension to encode additional symbols. In other words, the vicinity of the poles enables multi-radius constellations in which distinct states are separated primarily along the radial coordinate rather than angularly. Since the radius corresponds directly to optical power, this strategy introduces a controlled energy degree of freedom that can be optimized under linear-regime constraints of the PIP platform.

In this context, inspired by probabilistic constellation shaping techniques employed in digital coherent communication systems [34], the location of the states can be adapted according to their probability of occurrence, such that higher-probability states are placed in regions of lower noise over the GBS, see Fig.~\ref{fig:fig5}(c). In contrast to conventional digital approaches -- typically driven by symbol energy -- the APC framework enables probability shaping based on the spatial distribution of noise, including -- but not limited to -- spheres of different radius.

Therefore, the combination of (\emph{i}) non-uniform angular state distribution and (\emph{ii}) radial multiplexing near low-angular-sensitivity regions, together with probability shaping of the constellation states, constitutes a principled method to design noise- and energy-adapted analog constellations. In this sense, the present noise formalism complements the geometric design criteria of API [25] by introducing statistical weighting into the constellation optimization problem.

\noindent\textbf{Application to photonic neuromorphic computing.} The noise model developed in this work can be naturally extended to optical neuromorphic computing systems [12,22]. As shown in ref. [24], APC can naturally describe neuromorphic architectures, since an anbit may be interpreted as two neuromorphic units of information. Under this perspective, an artificial neural network with $N$ inputs and $N$ outputs can be reformulated as a multi-anbit APC system with $N/2$ input anbits and $N/2$ output anbits, each one represented in a different GBS. Consequently, the system noise can be described as a collection of noise maps distributed across $N/2$ GBSs. This geometric picture enables a direct extension of the noise-aware design principles introduced in this work: the input signals of the neural network can be grouped pairwise and encoded as anbit states, whose spatial location in each GBS is optimized according to the corresponding noise map at the receiver. In this way, rather than solely relying on training-based robustness, neuromorphic systems can leverage information-encoding strategies that minimize noise-induced errors at the hardware level. This approach establishes a direct link between noise physics and information encoding techniques in photonic neural networks, opening new avenues for noise-aware neuromorphic system design.

\noindent\textbf{Future directions for noise modeling in APC/API.} Several research directions emerge from this work. First, extending the analysis to feedback architectures (sequential APC systems) is essential, as anbit states circulate in closed loops and may undergo recursive noise accumulation. Understanding the impact of loop delay, and phase conditions on noise statistics is therefore critical to ensure computational stability. Second, the noise formalism should be generalized to multi-anbit gates to support large-scale systems [24,25]. In this context, correlations between EDF fluctuations may arise, requiring a covariance-based framework in the multi-dimensional GBS, potentially complemented by information-theoretic tools to fully characterize noise-induced dependencies. Third, more advanced O/E converters should be investigated. While differential schemes capture fundamental noise mechanisms, phase-diversity receivers could provide full access to the GBS and enable improved noise analysis [23,35]. Finally, incorporating deterministic perturbations from fabrication imperfections into the framework would enable calibration and compensation strategies, paving the way toward robust, scalable, and fault-tolerant APC systems.

In summary, this work provides a comprehensive statistical description of noise within the APC and API paradigms, bridging physical noise sources, geometric state representation, and information-theoretic design. By integrating noise-optimized modeling into the constellation and receiver design process, APC advances toward a scalable and robust optical computing paradigm fully aligned with the technological capabilities of PIP.

% --- Back matter ---
\section*{Author contributions}

Raúl López March developed the theoretical framework and carried out the numerical simulations and experimental measurements. Andrés Macho Ortiz contributed to the refinement of the theory. Andrés Macho Ortiz and José Capmany supervised the work. All authors contributed to the preparation of the manuscript.

\section*{Disclosures}

The authors have declared no conflict of interest.

\section*{Code and Data Availability}

The code and data that support the findings of this study are available from the corresponding authors upon reasonable request.

\section*{Acknowledgements}

This work was supported by ERC-ADG-2022-101097092 ANBIT, ERC-POC-2025-1 101241773 TRANSBIT MESH, GVA PROMETEO 2021/015 research excellency award, Fundación BBVA Programa de Investigación Fundamentos 2024 API project, Ministerio de Ciencia y Universidades Plan Complementario de Comunicación Cuántica projects QUANTUMABLE-1 and QUANTUMABLE-2, and HUB de Comunicaciones Cuánticas.

% --- References ---
\section*{References}

\begin{enumerate}[label={[\arabic*]},leftmargin=*,itemsep=2pt]

\item R. K. Cavin, P. Lugli, and V. V. Zhirnov, ``Science and engineering beyond Moore's law,'' Proceedings of the IEEE \textbf{100}, 1720 (2012).

\item I. L. Markov, ``Limits on fundamental limits to computation,'' Nature \textbf{512}, 147 (2014).

\item J. Shalf, ``The future of computing beyond Moore's law,'' Philosophical Transactions of the Royal Society A \textbf{378}, 20190061 (2020).

\item F. Zangeneh-Nejad, D. L. Sounas, A. Alù, and R. Fleury, ``Analogue computing with metamaterials,'' Nature Reviews Materials \textbf{6}, 207 (2021).

\item G. Wetzstein, A. Ozcan, S. Gigan, S. Fan, D. Englund, M. Soljačić, C. Denz, D. A. B. Miller, and D. Psaltis, ``Inference in artificial intelligence with deep optics and photonics,'' Nature \textbf{588}, 39 (2020).

\item S. Wang, W. Li, C. Yang, Y. Lu, W. Zhang, X. Li, J. Tang, B. Gao, H. Wu, and H. Qian, ``In-memory analog solution of compressed sensing recovery in one step,'' Science Advances \textbf{13}, 9 (2023).

\item S. Hua, E. Divita, S. Yu, B. Peng, C. Roques-Carmes, Z. Su, Z. Chen, Y. Bai, J. Zou, Y. Zhu, et al., ``An integrated large-scale photonic accelerator with ultralow latency,'' Nature \textbf{640}, 361 (2025).

\item M. Chen, Y. Lin, M. Duan, R. Midya, W. Zhang, P. Yan, J. J. Yang, and Q. Xia, ``Nanoelectronics-enabled reservoir computing hardware for real-time robotic controls,'' Science Advances \textbf{26}, 11 (2025).

\item X. X. Zhu, D. Tuia, L. Mou, G. S. Xia, L. Zhang, F. Xu, and F. Fraundorfer, ``Deep learning in remote sensing: a comprehensive review and list of resources,'' IEEE Geoscience and Remote Sensing Magazine \textbf{5}, 8 (2017).

\item C. V. Poulton, M. J. Byrd, P. Russo, B. Moss, O. Shatrovoy, M. Khandaker, and M. R. Watts, ``Coherent LiDAR with an 8,192-element optical phased array and driving laser,'' IEEE Journal of Selected Topics in Quantum Electronics \textbf{28}, 6100408 (2022).

\item L. Bombieri, Z. Zeng, R. Tricarico, R. Lin, S. Notarnicola, M. Cain, M. D. Lukin, and H. Pichler, ``Quantum adiabatic optimization with Rydberg arrays: localization phenomena and encoding strategies,'' PRX Quantum \textbf{6}, 020306 (2025).

\item B. J. Shastri, A. N. Tait, T. Ferreira de Lima, W. H. P. Pernice, H. Bhaskaran, C. D. Wright, and P. R. Prucnal, ``Photonics for artificial intelligence and neuromorphic computing,'' Nature Photonics \textbf{15}, 102 (2021).

\item J. Feldmann, N. Youngblood, M. Karpov, H. Gehring, X. Li, M. Stappers, M. Le Gallo, X. Fu, A. Lukashchuk, A. S. Raja, et al., ``Parallel convolutional processing using an integrated photonic tensor core,'' Nature \textbf{589}, 52 (2021).

\item X. Xu, M. Tan, B. Corcoran, J. Wu, A. Boes, T. G. Nguyen, S. T. Chu, B. E. Little, D. G. Hicks, R. Morandotti, A. Mitchell, and D. J. Moss, ``11 TOPS photonic convolutional accelerator for optical neural networks,'' Nature \textbf{589}, 44 (2021).

\item PsiQuantum Team, ``A manufacturable platform for photonic quantum computing,'' Nature \textbf{641}, 876 (2025).

\item W. Bogaerts, D. Pérez, J. Capmany, D. A. B. Miller, J. Poon, D. Englund, F. Morichetti, and A. Melloni, ``Programmable photonic circuits,'' Nature \textbf{586}, 207 (2020).

\item D. Pérez, I. Gasulla, P. DasMahapatra, and J. Capmany, ``Principles, fundamentals, and applications of programmable integrated photonics,'' Advances in Optics and Photonics \textbf{12}, 709 (2020).

\item J. Capmany and D. Pérez, \emph{Programmable Integrated Photonics}. (Oxford University Press, Oxford, 2020).

\item Z. Zhu, G. R. Kumar, K. Chen, A. Varri, and H. Saltık, ``Coherent general-purpose photonic matrix processor,'' ACS Photonics \textbf{11}, 1189 (2024).

\item M. Chen, Y. Tang, J. Shi, J. Shi, M. Tan, X. Xu, and J. Wu, ``I/O-efficient iterative matrix inversion with photonic integrated circuits,'' Nature Communications \textbf{15}, 5926 (2024).

\item J. Bao, Z. Fu, T. Pramanik, J. Mao, Y. Chi, Y. Cao, C. Zhai, Y. Mao, T. Dai, X. Chen, et al., ``Very-large-scale integrated quantum graph photonics,'' Nature Photonics \textbf{17}, 573 (2023).

\item T. Fu, Y. Zang, Y. Huang, Z. Du, H. Huang, C. Hu, M. Chen, S. Yang, and H. Chen, ``Optical neural networks: progress and challenges,'' Light: Science \& Applications \textbf{13}, 263 (2024).

\item Z. Xu, B. Tang, X. Zhang, J. F. Leong, J. Pan, S. Hooda, E. Zamburg, and A. V.-Y. Thean, ``Reconfigurable nonlinear photonic activation function for photonic neural network based on non-volatile opto-resistive RAM switch,'' Light: Science \& Applications \textbf{11}, 288 (2022).

\item A. Macho-Ortiz, D. Pérez-López, J. Azaña, and J. Capmany, ``Analog programmable-photonic computation,'' Laser \& Photonics Reviews \textbf{17}, 2200360 (2023).

\item A. Macho-Ortiz, R. López-March, P. Martínez-Carrasco, F. J. Fraile-Peláez, and J. Capmany, ``Analog programmable-photonic information,'' Advanced Photonics \textbf{8}, 036002 (2026).

\item E. Desurvire, \emph{Classical and Quantum Information Theory: An Introduction for the Telecom Scientist}. (Cambridge University Press, Cambridge, 2009).

\item J. G. Proakis and M. Salehi, \emph{Digital Communications}. (McGraw-Hill, New York, 2008).

\item K. Petermann, \emph{Laser Diode Modulation and Noise}. (Kluwer Academic Publishers, London, 1988).

\item G. P. Agrawal, \emph{Fiber-Optic Communication Systems}. (Wiley, New Jersey, 2010).

\item M. Jacques, A. Samani, E. E.-Fiky, D. Patel, Z. Xing, and D. V. Plant, ``Optimization of thermo-optic phase-shifter design and mitigation of thermal crosstalk on the SOI platform,'' Optics Express \textbf{27}, 10456 (2019).

\item E. Desurvire, \emph{Erbium-Doped Fiber Amplifiers: Principles and Applications}. (Wiley-Interscience, New York, 2002).

\item A. Macho, D. Pérez, and J. Capmany, ``Optical implementation of 2$\times$2 universal unitary matrix transformations,'' Laser \& Photonics Reviews \textbf{15}, 2000473 (2021).

\item F. A. Morrison, \emph{Uncertainty Analysis for Engineers and Scientists: A Practical Guide}. (Cambridge University Press, Cambridge, 2021).

\item J. Cho and P. J. Winzer, ``Probabilistic constellation shaping for optical fiber communications,'' Journal of Lightwave Technology \textbf{37}, 1590 (2019).

\item H. Tan, H. Chen, J. Wang, Y. Wang, and X. Zhang, ``C-band optical 90-degree hybrid using thin film lithium niobate,'' Optics Letters \textbf{48}, 1946 (2023).

\end{enumerate}

% ==========================================================
% SUPPLEMENTARY INFORMATION
% ==========================================================

\newpage

% Reset counters and relabel for supplementary
\setcounter{section}{0}
\setcounter{equation}{0}
\setcounter{figure}{0}
\setcounter{table}{0}
\renewcommand{\thesection}{\arabic{section}}
\renewcommand{\theequation}{S\arabic{section}.\arabic{equation}}
\renewcommand{\thefigure}{S\arabic{figure}}
\makeatletter
\@addtoreset{equation}{section}
\makeatother

\begin{center}
    {\Large \textbf{Supplementary information:}}\\[0.3em]
    {\Large \textbf{Noise in analog programmable-photonic computation}}\\[1em]
    Raúl López-March$^{a}$, Andrés Macho-Ortiz$^{a,*}$, Francisco Javier Fraile-Peláez$^{b}$, and José Capmany$^{a,c,*}$\\[0.5em]
    {\small $^{a}$ITEAM Research Institute, Universitat Politècnica de València, Valencia, 46022, Spain}\\
    {\small $^{b}$Dept. Teoría de la Señal y Comunicaciones, Universidad de Vigo E.I. Telecomunicación, Campus Universitario, E-36202 Vigo (Pontevedra), Spain}\\
    {\small $^{c}$iPronics, Programmable Photonics, S.L, Camino de Vera s/n, Valencia, 46022, Spain}\\[0.3em]
    {\small $^{*}$Corresponding authors: \href{mailto:amachor@iteam.upv.es}{amachor@iteam.upv.es}, \href{mailto:jcapmany@iteam.upv.es}{jcapmany@iteam.upv.es}}
\end{center}

\vspace{1em}

In this Supplementary Information, we provide a detailed account of the theory underpinning the noise analysis in analog programmable-photonic computing (APC), together with additional information on the numerical simulations, experimental measurements, and methods used in this work. Equations and figures are labeled with the prefix `S' to distinguish them from those in the main text.

\section{Differential O/E converters: photocurrents}

\begin{figure}[h!]
    \centering
    \includegraphics[width=0.8\textwidth]{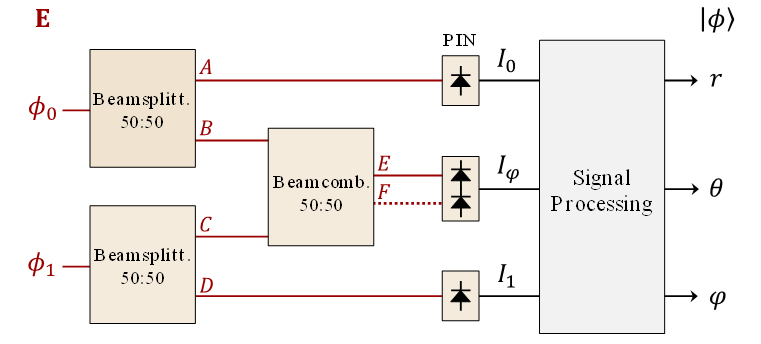}
    \caption{\textbf{Differential O/E converters.} Common hardware scheme representing both unbalanced and balanced designs. The unbalanced configuration does not include the dashed line. Here, we also depict the complex envelopes generated at the outputs of the optical devices, denoted by capital letters \emph{A}, \emph{B}, \emph{C}, etc.}
    \label{fig:figS1}
\end{figure}

Figure~\ref{fig:figS1} depicts the schematic circuitry for both unbalanced and balanced differential opto-electrical (O/E) converters, detailing the complex envelopes generated at the outputs of the optical devices, subsequently employed in the mathematical calculations.

While the theoretical analysis of the unbalanced receiver is reported in ref. [1], the derivation of the photocurrents for the balanced design still remains to be carried out. Both receivers can be commonly analyzed using the scheme shown in Fig.~\ref{fig:figS1}. The electric field $\mathbf{E}$ at the input of the O/E converter is composed of two complex envelopes (or wave packets) $\phi_0$ and $\phi_1$ with a phase delay $\varphi=\arg\phi_1-\arg\phi_0$ between them. Here, the functionality of both O/E converters is to retrieve the \emph{effective degrees of freedom} (EDFs) $r$, $\theta$, and $\varphi$ that constitute the received state of the analog bit (anbit) $|\phi\rangle = \phi_0|0\rangle + e^{j\varphi}\phi_1|1\rangle \equiv r\cos(\theta/2)|0\rangle + e^{j\varphi}\sin(\theta/2)|1\rangle$ via the photocurrents $I_0$, $I_1$, and $I_\varphi$. These photocurrents can be derived from the \emph{transfer matrices} of the 50:50 beam splitters (BS), implemented via Y-junctions [2], and the 50:50 beam combiner (BC), realized using a multi-mode interferometer [3]:
\begin{equation}
T_{\mathrm{BS}} = \frac{e^{j\delta}}{\sqrt{2}}\begin{pmatrix}1 \\ 1\end{pmatrix}, \qquad T_{\mathrm{BC}} = \frac{e^{j\delta}}{\sqrt{2}}\begin{pmatrix}1 & j \\ j & 1\end{pmatrix},
\label{eq:S1.1}
\end{equation}
where the global phase $\delta$ is an unknown design parameter that depends on the manufacturing process of the devices. Therefore, this global phase should be assumed to be different for each device in the receiver.

Specifically, the envelopes at the output of the 50:50 BSs are:
\begin{equation}
\begin{pmatrix}A \\ B\end{pmatrix} = \frac{e^{j\delta_0}}{\sqrt{2}}\begin{pmatrix}1 \\ 1\end{pmatrix}\phi_0 = \begin{pmatrix}\frac{e^{j\delta_0}}{\sqrt{2}}\phi_0 \\ \frac{e^{j\delta_0}}{\sqrt{2}}\phi_0\end{pmatrix},
\label{eq:S1.2}
\end{equation}
\begin{equation}
\begin{pmatrix}C \\ D\end{pmatrix} = \frac{e^{j\delta_1}}{\sqrt{2}}\begin{pmatrix}1 \\ 1\end{pmatrix}\phi_1 = \begin{pmatrix}\frac{e^{j\delta_1}}{\sqrt{2}}\phi_1 \\ \frac{e^{j\delta_1}}{\sqrt{2}}\phi_1\end{pmatrix},
\label{eq:S1.3}
\end{equation}
and the envelopes at the output of the 50:50 BC are:
\begin{equation}
\begin{pmatrix}E \\ F\end{pmatrix} = \frac{e^{j\delta_\varphi}}{\sqrt{2}}\begin{pmatrix}1 & j \\ j & 1\end{pmatrix}\begin{pmatrix}B \\ C\end{pmatrix} = \frac{e^{j\delta_\varphi}}{2}\begin{pmatrix}e^{j\delta_0}\phi_0 + j e^{j\delta_1}\phi_1 \\ j e^{j\delta_0}\phi_0 + e^{j\delta_1}\phi_1\end{pmatrix}.
\label{eq:S1.4}
\end{equation}

Once we have obtained the expression of the complex envelopes at the input of the photodiodes, the corresponding photocurrents can be derived. The photocurrents $I_0$ and $I_1$ are found to be the same in both receivers:
\begin{equation}
I_0 = \mathcal{R}|A|^2 = \tfrac{1}{2}\mathcal{R}|\phi_0|^2, \qquad I_1 = \mathcal{R}|D|^2 = \tfrac{1}{2}\mathcal{R}|\phi_1|^2,
\label{eq:S1.5}
\end{equation}
where $\mathcal{R}$ is the responsivity of the photodiode. Nevertheless, the photocurrent $I_\varphi$ differs between the unbalanced and balanced designs. In particular, in the unbalanced converter, we find that:
\begin{equation}
I_\varphi = \mathcal{R}|E|^2 = \tfrac{1}{4}\mathcal{R}\left[|\phi_0|^2 + |\phi_1|^2 - 2|\phi_0||\phi_1|\sin(\delta_1-\delta_0+\varphi)\right],
\label{eq:S1.6}
\end{equation}
whereas, in the balanced configuration, we obtain:
\begin{equation}
I_\varphi = \mathcal{R}\left(|E|^2 - |F|^2\right) = -\mathcal{R}|\phi_0||\phi_1|\sin(\delta_1-\delta_0+\varphi).
\label{eq:S1.7}
\end{equation}

In this work, we assume $\delta_0=\delta_1$ in order to separate noise-induced computational errors from those arising from hardware imperfections. The case $\delta_0\neq\delta_1$, together with additional hardware non-idealities, will be addressed in forthcoming APC contributions, as commented in the Discussion section of the main text.

Finally, the EDFs $r$, $\theta$, and $\varphi$ can be recovered from the photocurrents using Eqs.~(\ref{eq:S1.5})-(\ref{eq:S1.7}) and considering that $|\phi_0|=r\cos(\theta/2)$ and $|\phi_1|=r\sin(\theta/2)$. Thus, the radius of the Generalized Bloch Sphere (GBS) is found to be:
\begin{equation}
r = \sqrt{\frac{2(I_0+I_1)}{\mathcal{R}}},
\label{eq:S1.8}
\end{equation}
the elevation angle is:
\begin{equation}
\theta = 2\arctan\sqrt{\frac{I_1}{I_0}},
\label{eq:S1.9}
\end{equation}
and the azimuthal angle (or differential phase) is:
\begin{equation}
\varphi_{\mathrm{(unbalanced)}} = \arcsin\left(\frac{I_0+I_1-2I_\varphi}{2\sqrt{I_0 I_1}}\right), \qquad \varphi_{\mathrm{(balanced)}} = \arcsin\left(\frac{-I_\varphi}{2\sqrt{I_0 I_1}}\right).
\label{eq:S1.10}
\end{equation}

Note that the above expressions are Eqs.~(2)-(5) of the paper.

\section{First-order error propagation theory}

First-order error propagation theory (EPT) provides a linearized description of how uncertainties in a set of input random variables $\mathbf{x}=(x_1,x_2,\ldots,x_n)$ map to the uncertainty of a derived quantity $y$ [4]. Let us describe in detail how first-order EPT operates. Consider a differentiable scalar function of real random variables $y=f(\mathbf{x})$, where each input random variable $x_i$ has a mean $\overline{x_i}$, such that the vector mean is $\overline{\mathbf{x}}=(\overline{x_1},\overline{x_2},\ldots,\overline{x_n})$. When the fluctuations $\delta x_i \coloneqq x_i - \overline{x_i}$ are sufficiently small to guarantee that the first-order differential:
\begin{equation}
\mathrm{d}y = \mathrm{d}f(\overline{\mathbf{x}})(\delta\mathbf{x}) = \sum_{i=1}^{n} \frac{\partial f(\overline{\mathbf{x}})}{\partial x_i}\delta x_i,
\label{eq:S2.1}
\end{equation}
dominates over higher-order differentials, first-order EPT states that the fluctuations $\delta y \coloneqq y - f(\overline{\mathbf{x}})$ can be approximated as $\delta y \approx \mathrm{d}y$. Consequently, the fluctuations $\delta y$ can be calculated as:
\begin{equation}
\delta y \approx \mathrm{d}f(\overline{\mathbf{x}})(\delta\mathbf{x}) = \sum_{i=1}^{n} \frac{\partial f(\overline{\mathbf{x}})}{\partial x_i}\delta x_i.
\label{eq:S2.2}
\end{equation}

Thus, the validity limits of first-order EPT coincide with those of the first-order Taylor series expansion of the function $f$ around the point $\mathbf{x}=\overline{\mathbf{x}}$.

Next, considering that $\delta x_1, \delta x_2, \ldots, \delta x_n$ are zero-mean random variables, we infer that $\delta y$ is also a \emph{zero-mean} random variable by applying the expectation operator ($\mathbb{E}$) to Eq.~(\ref{eq:S2.2}), that is, $\mathbb{E}(\delta y)=0$. Accordingly, the variance of $y$ is equal to the variance of $\delta y$:
\begin{align}
\mathrm{Var}(y) &= \mathbb{E}(y^2) - \left[\mathbb{E}(y)\right]^2 \nonumber \\
&= \mathbb{E}\left[\left(f(\overline{\mathbf{x}})+\delta y\right)^2\right] - \left[\mathbb{E}\left(f(\overline{\mathbf{x}})+\delta y\right)\right]^2 \nonumber \\
&= \mathbb{E}\left[f(\overline{\mathbf{x}})^2 + (\delta y)^2 + 2f(\overline{\mathbf{x}})\delta y\right] - f(\overline{\mathbf{x}})^2 \nonumber \\
&= f(\overline{\mathbf{x}})^2 + \mathbb{E}\left[(\delta y)^2\right] + 2f(\overline{\mathbf{x}})\,\mathbb{E}(\delta y) - f(\overline{\mathbf{x}})^2 \nonumber \\
&= \mathbb{E}\left[(\delta y)^2\right] \equiv \mathrm{Var}(\delta y).
\label{eq:S2.3}
\end{align}

Consequently, the variance of $y$ is obtained by calculating the second moment of $\delta y$, which yields the standard EPT expression:
\begin{align}
\mathrm{Var}(y) &= \mathrm{Var}(\delta y) = \mathbb{E}\left[(\delta y)^2\right] \approx \mathbb{E}\left[\left(\sum_i \frac{\partial f(\overline{\mathbf{x}})}{\partial x_i}\delta x_i\right)^2\right] = \mathbb{E}\left[\sum_{i,j}\frac{\partial f(\overline{\mathbf{x}})}{\partial x_i}\frac{\partial f(\overline{\mathbf{x}})}{\partial x_j}\delta x_i\delta x_j\right] \nonumber \\
&= \mathbb{E}\left[\sum_i \left(\frac{\partial f(\overline{\mathbf{x}})}{\partial x_i}\right)^2 (\delta x_i)^2\right] + \mathbb{E}\left[\sum_{i\neq j}\frac{\partial f(\overline{\mathbf{x}})}{\partial x_i}\frac{\partial f(\overline{\mathbf{x}})}{\partial x_j}\delta x_i\delta x_j\right] \nonumber \\
&= \sum_i \left(\frac{\partial f(\overline{\mathbf{x}})}{\partial x_i}\right)^2 \mathbb{E}\left[(\delta x_i)^2\right] + \sum_{i\neq j} \frac{\partial f(\overline{\mathbf{x}})}{\partial x_i}\frac{\partial f(\overline{\mathbf{x}})}{\partial x_j}\mathbb{E}(\delta x_i \delta x_j) \nonumber \\
&\equiv \sum_i \left(\frac{\partial f(\overline{\mathbf{x}})}{\partial x_i}\right)^2 \mathrm{Var}(\delta x_i) + 2\sum_{i<j} \frac{\partial f(\overline{\mathbf{x}})}{\partial x_i}\frac{\partial f(\overline{\mathbf{x}})}{\partial x_j}\mathrm{Cov}(\delta x_i, \delta x_j),
\label{eq:S2.4}
\end{align}
where ``Cov'' is the covariance. The factor 2 and the condition $i<j$ in the dummy indices are included to account for the symmetry of the covariance, since $\mathrm{Cov}(\delta x_i,\delta x_j)=\mathrm{Cov}(\delta x_j,\delta x_i)$.

In our problem, we have three different derived quantities, the EDFs of the received state, which depend on the photocurrents. Thus, the derived quantity $y$ corresponds to one of the specific EDFs $r$, $\theta$, or $\varphi$; whereas the input vector $\mathbf{x}$ should be identified with the 3-tuple $(I_0, I_1, I_\varphi)$. Therefore, calculating the partial derivatives in Eq.~(\ref{eq:S2.4}), it is straightforward to derive the closed-form expressions of $\mathrm{Var}(r)$, $\mathrm{Var}(\theta)$, and $\mathrm{Var}(\varphi)$.\\

\noindent\textbf{Radius}

We know that:
\begin{equation}
r = \sqrt{\frac{2(I_0+I_1)}{\mathcal{R}}} \equiv f(I_0, I_1, I_\varphi).
\label{eq:S2.5}
\end{equation}
The partial derivatives required in Eq.~(\ref{eq:S2.4}) are:
\begin{equation}
\frac{\partial f(\overline{I_0},\overline{I_1},\overline{I_\varphi})}{\partial I_0} = \frac{\partial f(\overline{I_0},\overline{I_1},\overline{I_\varphi})}{\partial I_1} = \left[2\mathcal{R}(\overline{I_0}+\overline{I_1})\right]^{-1/2},
\label{eq:S2.6}
\end{equation}
and $\partial f(\overline{I_0},\overline{I_1},\overline{I_\varphi})/\partial I_\varphi = 0$. Consequently, the variance of the radius can be approximated as:
\begin{equation}
\mathrm{Var}(r) \approx \frac{1}{2\mathcal{R}(\overline{I_0}+\overline{I_1})}\left[\mathrm{Var}(\delta I_0)+\mathrm{Var}(\delta I_1)\right] + \text{covariance terms}.
\label{eq:S2.7}
\end{equation}

Furthermore, noting that $\overline{I_0}$ and $\overline{I_1}$ can be written as:
\begin{equation}
\overline{I_0} = \tfrac{1}{2}\mathcal{R}\overline{r}^2\cos^2\frac{\overline{\theta}}{2}, \qquad \overline{I_1} = \tfrac{1}{2}\mathcal{R}\overline{r}^2\sin^2\frac{\overline{\theta}}{2},
\label{eq:S2.8}
\end{equation}
where $\overline{r}=\mathbb{E}(r)$ and $\overline{\theta}=\mathbb{E}(\theta)$, it follows that Eq.~(\ref{eq:S2.7}) reduces to:
\begin{equation}
\mathrm{Var}(r) \approx \frac{1}{\mathcal{R}^2\overline{r}^2}\left[\mathrm{Var}(\delta I_0)+\mathrm{Var}(\delta I_1)\right] + \text{covariance terms},
\label{eq:S2.9}
\end{equation}
which is consistent with Eq.~(6) of the main text when the covariance terms -- discussed at the end of this section -- are omitted.\\

\noindent\textbf{Elevation angle}

The elevation angle is given by the expression:
\begin{equation}
\theta = 2\arctan\sqrt{\frac{I_1}{I_0}} \equiv f(I_0, I_1, I_\varphi).
\label{eq:S2.10}
\end{equation}
The partial derivatives required in Eq.~(\ref{eq:S2.4}) are:
\begin{equation}
\frac{\partial f(\overline{I_0},\overline{I_1},\overline{I_\varphi})}{\partial I_0} = -\frac{1}{\overline{I_0}+\overline{I_1}}\sqrt{\frac{\overline{I_1}}{\overline{I_0}}},
\label{eq:S2.11}
\end{equation}
\begin{equation}
\frac{\partial f(\overline{I_0},\overline{I_1},\overline{I_\varphi})}{\partial I_1} = +\frac{1}{\overline{I_0}+\overline{I_1}}\sqrt{\frac{\overline{I_0}}{\overline{I_1}}},
\label{eq:S2.12}
\end{equation}
and $\partial f(\overline{I_0},\overline{I_1},\overline{I_\varphi})/\partial I_\varphi = 0$. Hence, the variance of the elevation angle can be approximated as:
\begin{equation}
\mathrm{Var}(\theta) \approx \frac{1}{(\overline{I_0}+\overline{I_1})^2}\left[\frac{\overline{I_1}}{\overline{I_0}}\mathrm{Var}(\delta I_0) + \frac{\overline{I_0}}{\overline{I_1}}\mathrm{Var}(\delta I_1)\right] + \text{covariance terms}.
\label{eq:S2.13}
\end{equation}

Next, using Eq.~(\ref{eq:S2.8}), the above expression becomes:
\begin{equation}
\mathrm{Var}(\theta) \approx \frac{4}{\mathcal{R}^2\overline{r}^4}\left[\tan^2\frac{\overline{\theta}}{2}\mathrm{Var}(\delta I_0) + \cot^2\frac{\overline{\theta}}{2}\mathrm{Var}(\delta I_1)\right] + \text{covariance terms},
\label{eq:S2.14}
\end{equation}
which is Eq.~(7) of the paper when the covariance terms are neglected.\\

\noindent\textbf{Azimuthal angle (or differential phase): unbalanced design}

In the unbalanced differential O/E converter, the differential phase is:
\begin{equation}
\varphi = \arcsin\left(\frac{I_0+I_1-2I_\varphi}{2\sqrt{I_0 I_1}}\right) \equiv f(I_0, I_1, I_\varphi).
\label{eq:S2.15}
\end{equation}
The calculation of the partial derivatives required in Eq.~(\ref{eq:S2.4}) simplifies by introducing the auxiliary variable $\beta \coloneqq (I_0+I_1-2I_\varphi)/(2\sqrt{I_0 I_1}) \equiv \sin\varphi$. In this way:
\begin{equation}
\frac{\partial \varphi}{\partial I_k} = \frac{1}{\sqrt{1-\beta^2}}\frac{\partial \beta}{\partial I_k} = \frac{1}{\cos\varphi}\frac{\partial \beta}{\partial I_k}, \quad k\in\{0,1,\varphi\},
\label{eq:S2.16}
\end{equation}
and:
\begin{equation}
\mathrm{Var}(\varphi) \approx \frac{1}{\cos^2\overline{\varphi}}\sum_k\left(\frac{\partial\overline{\beta}}{\partial I_k}\right)^2\mathrm{Var}(\delta I_k) + \text{covariance terms}.
\label{eq:S2.17}
\end{equation}

Hence, the derivation of $\mathrm{Var}(\varphi)$ reduces to calculating $\partial\beta/\partial I_k$, evaluated at the ideal values of the photocurrents $(\partial\overline{\beta}/\partial I_k)$. After some algebraic manipulations and using Eq.~(\ref{eq:S2.8}), we obtain:
\begin{equation}
\frac{\partial\overline{\beta}}{\partial I_0} = \frac{1-\sin\overline{\varphi}\sqrt{\overline{I_1}/\overline{I_0}}}{2\sqrt{\overline{I_0}\,\overline{I_1}}} = \frac{1-\sin\overline{\varphi}\tan(\overline{\theta}/2)}{\tfrac{1}{2}\mathcal{R}\overline{r}^2\sin\overline{\theta}},
\label{eq:S2.18}
\end{equation}
\begin{equation}
\frac{\partial\overline{\beta}}{\partial I_1} = \frac{1-\sin\overline{\varphi}\sqrt{\overline{I_0}/\overline{I_1}}}{2\sqrt{\overline{I_0}\,\overline{I_1}}} = \frac{1-\sin\overline{\varphi}\cot(\overline{\theta}/2)}{\tfrac{1}{2}\mathcal{R}\overline{r}^2\sin\overline{\theta}},
\label{eq:S2.19}
\end{equation}
\begin{equation}
\frac{\partial\overline{\beta}}{\partial I_\varphi} = -\frac{1}{\sqrt{\overline{I_0}\,\overline{I_1}}} = -\frac{4}{\mathcal{R}\overline{r}^2\sin\overline{\theta}}.
\label{eq:S2.20}
\end{equation}

Accordingly, Eq.~(\ref{eq:S2.17}) becomes:
\begin{multline}
\mathrm{Var}(\varphi) \approx \frac{1}{\cos^2\overline{\varphi}}\Bigg[\left(\frac{1-\sin\overline{\varphi}\tan(\overline{\theta}/2)}{\tfrac{1}{2}\mathcal{R}\overline{r}^2\sin\overline{\theta}}\right)^2\mathrm{Var}(\delta I_0) + \left(\frac{1-\sin\overline{\varphi}\cot(\overline{\theta}/2)}{\tfrac{1}{2}\mathcal{R}\overline{r}^2\sin\overline{\theta}}\right)^2\mathrm{Var}(\delta I_1) \\
+ \left(\frac{4}{\mathcal{R}\overline{r}^2\sin\overline{\theta}}\right)^2\mathrm{Var}(\delta I_\varphi)\Bigg] + \text{covariance terms} \\
= \frac{4}{\mathcal{R}^2\overline{r}^4\sin^2\overline{\theta}\cos^2\overline{\varphi}}\Bigg[\left(1-\sin\overline{\varphi}\tan\frac{\overline{\theta}}{2}\right)^2\mathrm{Var}(\delta I_0) + \left(1-\sin\overline{\varphi}\cot\frac{\overline{\theta}}{2}\right)^2\mathrm{Var}(\delta I_1) \\
+ 4\,\mathrm{Var}(\delta I_\varphi)\Bigg] + \text{covariance terms},
\label{eq:S2.21}
\end{multline}
which reduces to Eq.~(8) of the main text when the covariance terms are omitted.\\

\noindent\textbf{Azimuthal angle (or differential phase): balanced design}

In the balanced differential O/E converter, the differential phase is:
\begin{equation}
\varphi = \arcsin\left(\frac{-I_\varphi}{2\sqrt{I_0 I_1}}\right) \equiv f(I_0, I_1, I_\varphi).
\label{eq:S2.22}
\end{equation}
Proceeding in a similar manner to the unbalanced case, the calculation of the partial derivatives required in Eq.~(\ref{eq:S2.4}) simplifies by introducing the auxiliary variable $\beta \coloneqq -I_\varphi/(2\sqrt{I_0 I_1}) \equiv \sin\varphi$. Thus:
\begin{equation}
\frac{\partial \varphi}{\partial I_k} = \frac{1}{\sqrt{1-\beta^2}}\frac{\partial \beta}{\partial I_k} = \frac{1}{\cos\varphi}\frac{\partial \beta}{\partial I_k}, \quad k\in\{0,1,\varphi\},
\label{eq:S2.23}
\end{equation}
and:
\begin{equation}
\mathrm{Var}(\varphi) \approx \frac{1}{\cos^2\overline{\varphi}}\sum_k\left(\frac{\partial\overline{\beta}}{\partial I_k}\right)^2\mathrm{Var}(\delta I_k) + \text{covariance terms}.
\label{eq:S2.24}
\end{equation}

Therefore, the derivation of $\mathrm{Var}(\varphi)$ reduces to calculating $\partial\beta/\partial I_k$, evaluated at the ideal values of the photocurrents $(\partial\overline{\beta}/\partial I_k)$. After some algebraic manipulations and using Eq.~(\ref{eq:S2.8}), we obtain:
\begin{equation}
\frac{\partial\overline{\beta}}{\partial I_0} = -\frac{\sin\overline{\varphi}}{2\overline{I_0}} = -\frac{\sin\overline{\varphi}}{\mathcal{R}\overline{r}^2\cos^2(\overline{\theta}/2)},
\label{eq:S2.25}
\end{equation}
\begin{equation}
\frac{\partial\overline{\beta}}{\partial I_1} = -\frac{\sin\overline{\varphi}}{2\overline{I_1}} = -\frac{\sin\overline{\varphi}}{\mathcal{R}\overline{r}^2\sin^2(\overline{\theta}/2)},
\label{eq:S2.26}
\end{equation}
\begin{equation}
\frac{\partial\overline{\beta}}{\partial I_\varphi} = -\frac{1}{2\sqrt{\overline{I_0}\,\overline{I_1}}} = -\frac{2}{\mathcal{R}\overline{r}^2\sin\overline{\theta}}.
\label{eq:S2.27}
\end{equation}

Finally, Eq.~(\ref{eq:S2.24}) becomes:
\begin{multline}
\mathrm{Var}(\varphi) \approx \frac{1}{\cos^2\overline{\varphi}}\Bigg[\left(\frac{\sin\overline{\varphi}}{\mathcal{R}\overline{r}^2\cos^2(\overline{\theta}/2)}\right)^2\mathrm{Var}(\delta I_0) + \left(\frac{\sin\overline{\varphi}}{\mathcal{R}\overline{r}^2\sin^2(\overline{\theta}/2)}\right)^2\mathrm{Var}(\delta I_1) \\
+ \left(\frac{2}{\mathcal{R}\overline{r}^2\sin\overline{\theta}}\right)^2\mathrm{Var}(\delta I_\varphi)\Bigg] + \text{covariance terms} \\
= \frac{1}{\mathcal{R}^2\overline{r}^4\cos^2\overline{\varphi}}\Bigg[\left(\frac{\sin\overline{\varphi}}{\cos^2(\overline{\theta}/2)}\right)^2\mathrm{Var}(\delta I_0) + \left(\frac{\sin\overline{\varphi}}{\sin^2(\overline{\theta}/2)}\right)^2\mathrm{Var}(\delta I_1) + \left(\frac{2}{\sin\overline{\theta}}\right)^2\mathrm{Var}(\delta I_\varphi)\Bigg] \\
+ \text{covariance terms},
\label{eq:S2.28}
\end{multline}
which reduces to Eq.~(9) of the main text when the covariance terms are omitted.\\

\noindent\textbf{Covariance terms: discussion}

\noindent{In} general, the covariance terms in Eqs.~(\ref{eq:S2.9}), (\ref{eq:S2.14}), (\ref{eq:S2.21}), and (\ref{eq:S2.28}) are non-zero. For instance, $\mathrm{Cov}(\delta I_0, \delta I_1) \neq 0$, as a linear relation between $\delta I_0$ and $\delta I_1$ is induced by the relative intensity noise (RIN) of the (common) laser generating both anbit amplitudes $\phi_0$ and $\phi_1$. Nevertheless, these covariance terms are neglected in Eqs.~(6)-(9) of the main text, as the resulting expressions are consistent with the numerical simulations and experimental measurements reported in the Results section. This, in turn, implies that, in the present work, the covariance terms are much smaller than the variance terms in the first-order EPT expression given by Eq.~(\ref{eq:S2.4}).

\section{Laser phase noise and phase shifter noise}

\noindent\textbf{Laser phase noise}

\noindent{Here}, we discuss the impact of the laser phase noise on the API system depicted in Fig.~1 of the paper and demonstrate that it is completely negligible. We start by considering the laser as a coherent light source with a time-varying stochastic phase fluctuation $\Delta\phi_L(t)$ [5]. Therefore, excluding the relative intensity noise of the laser, the phasor of the electric field at the laser output (and at the input of the space-anbit modulator of Fig.~1) can be assumed to be of the form [5]:
\begin{equation}
\mathbf{E}_L(t) = \mathbf{A}_0 e^{j(\omega_0 t + \Delta\phi_L(t))} \equiv \mathbf{E}_0(t) e^{j\Delta\phi_L(t)},
\label{eq:S3.1}
\end{equation}
where $\mathbf{E}_0$ is the ideal phasor. In noiseless conditions ($\Delta\phi_L=0$), the space-anbit modulator generates the anbit:
\begin{equation}
|\phi\rangle = r\left(\cos\frac{\theta}{2}|0\rangle + e^{j\varphi}\sin\frac{\theta}{2}|1\rangle\right),
\label{eq:S3.2}
\end{equation}
from $\mathbf{E}_0$. In real noisy conditions ($\Delta\phi_L\neq 0$), the space-anbit modulator generates the anbit:
\begin{equation}
|\widetilde{\phi}(t)\rangle = e^{j\Delta\phi_L(t)}\,r\left(\cos\frac{\theta}{2}|0\rangle + e^{j\varphi}\sin\frac{\theta}{2}|1\rangle\right),
\label{eq:S3.3}
\end{equation}
from $\mathbf{E}_L$. As seen, the laser phase noise introduces a global phase in the generated anbit. However, note that \emph{a global phase is not observable} in the GBS [1,6], i.e., the state $|\widetilde{\phi}(t)\rangle$ is the same as the state $|\phi\rangle$ in the GBS (the spherical coordinates are the same). Consequently, \emph{the laser phase noise does not constitute an observable perturbation} of the anbits in the GBS.\\

\noindent\textbf{Phase shifter noise}

\noindent{The} phase shift generated by a \emph{thermo-optic} phase shifter is governed by the equation [7,8]:
\begin{equation}
\Delta\phi = k_0 \cdot \Delta n_{\mathrm{eff}}(T_0) \cdot L = k_0 \cdot \frac{\Delta n_{\mathrm{eff}}(T_0)}{\Delta T}\cdot \Delta T \cdot L \approx k_0 \cdot n'_{\mathrm{eff}}(T_0)\cdot \Delta T \cdot L,
\label{eq:S3.4}
\end{equation}
where $k_0$ is the vacuum wavenumber, $\Delta n_{\mathrm{eff}}$ denotes the variation of the effective index ($n_{\mathrm{eff}}$) of the fundamental mode supported by the waveguide implementing the phase shifter, $T_0$ is the reference temperature, $\Delta T$ is the temperature excursion with respect to $T_0$, $n'_{\mathrm{eff}}(T_0)=\mathrm{d}n_{\mathrm{eff}}(T=T_0)/\mathrm{d}T$, and $L$ is the phase-shifter length.

A thermo-optic phase shifter introduces dynamic phase noise originating from undesired fluctuations in the electrical current driving the resistive heater, which produce unwanted temporal variations of the device temperature. In particular, if $\Delta T$ includes a dynamic perturbation $\Delta T_e(t)$ with respect to the desired value $\Delta T_0$ ($\Delta T = \Delta T_0 + \Delta T_e$), then we observe a dynamic additive phase error $\Delta\phi_e(t)$ with respect to the desired value $\Delta\phi_0$:
\begin{align}
\Delta\phi(t) &= k_0 \cdot n'_{\mathrm{eff}}(T_0) \cdot \left(\Delta T_0 + \Delta T_e(t)\right)\cdot L \nonumber \\
&= k_0 \cdot n'_{\mathrm{eff}}(T_0) \cdot \Delta T_0 \cdot L + k_0 \cdot n'_{\mathrm{eff}}(T_0) \cdot \Delta T_e(t) \cdot L \equiv \Delta\phi_0 + \Delta\phi_e(t).
\label{eq:S3.5}
\end{align}

Thus, the goal is to characterize analytically and experimentally the statistical distribution of the random process $\Delta\phi_e(t)$.

To this end, note that the relation between $\Delta T$ and the electrical current $I$ driving the heater is given by the expression [8]:
\begin{equation}
\Delta T = \frac{R_L}{G\cdot A}\cdot I^2,
\label{eq:S3.6}
\end{equation}
where $R_L$ is the load resistance modelling the phase shifter (in ohms), $G$ is the thermal conductance (in W$\cdot$K$^{-1}\cdot$m$^{-2}$), and $A$ is the cross-sectional area through which the heat flows (in m$^2$). Here, if the driving current $I$ exhibits a dynamic additive perturbation $I_e(t)$ with respect to the desired value $I_0$ ($I=I_0+I_e$), then Eq.~(\ref{eq:S3.6}) becomes:
\begin{equation}
\Delta T = \frac{R_L}{G\cdot A}\cdot\left(I_0+I_e(t)\right)^2 = \frac{R_L}{G\cdot A}\cdot\left(I_0^2 + 2I_0 I_e(t) + I_e^2(t)\right) \approx \frac{R_L}{G\cdot A}\cdot\left(I_0^2 + 2I_0 I_e(t)\right),
\label{eq:S3.7}
\end{equation}
where the term $I_e^2$ has been neglected under the assumption $\langle I_e(t) \rangle \ll I_0$. Hence, we can describe the dynamic temperature perturbation $\Delta T_e(t)$ of the form:
\begin{equation}
\Delta T_e(t) \approx \frac{2\cdot R_L \cdot I_0}{G\cdot A}\cdot I_e(t).
\label{eq:S3.8}
\end{equation}

Consequently, the random process $\Delta\phi_e(t)$ is found to be:
\begin{equation}
\Delta\phi_e(t) = k_0 \cdot n'_{\mathrm{eff}}(T_0) \cdot \Delta T_e(t) \cdot L \approx k_0 \cdot n'_{\mathrm{eff}}(T_0) \cdot L \cdot \frac{2\cdot R_L \cdot I_0}{G \cdot A}\cdot I_e(t) \equiv \alpha\cdot I_e(t),
\label{eq:S3.9}
\end{equation}
being $\alpha$ a real-positive constant. Assuming that $I_e(t)$ is a \emph{stationary} random process with probability density function (pdf) $f_{I_e}(i)$, then the pdf of $\Delta\phi_e(t)$ is:
\begin{equation}
f_{\Delta\phi_e}(\phi) = \frac{1}{\alpha}f_{I_e}\left(\frac{\phi}{\alpha}\right).
\label{eq:S3.10}
\end{equation}

\begin{figure}[h!]
    \centering
    \includegraphics[width=0.6\textwidth]{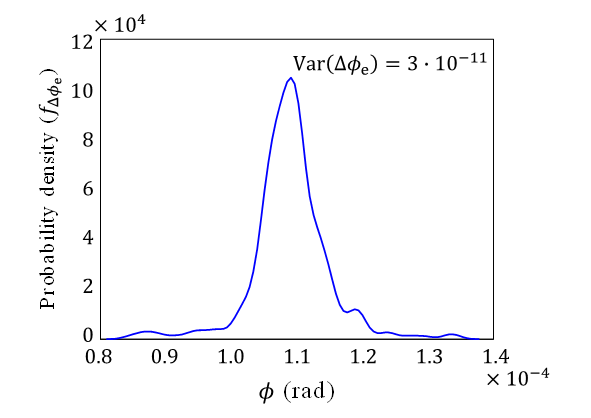}
    \caption{\textbf{Estimated phase-noise distribution in a thermo-optic phase shifter.} Probability density function (pdf) of the phase error $\Delta\phi_e$ induced by a thermo-optic phase shifter. The error phase is generated from undesired fluctuations in the electrical current driving the device. The pdf is estimated from the measured pdf of the current fluctuations by using Eq.~(\ref{eq:S3.10}).}
    \label{fig:figS2}
\end{figure}

By experimentally measuring $f_{I_e}$ for the current source driving the phase shifter, we can estimate $f_{\Delta\phi_e}$ using the above equation. Figure~\ref{fig:figS2} shows the estimated pdf $f_{\Delta\phi_e}$ for a driving current $I_0=1$~mA. The parameters required to estimate the constant $\alpha$ in Eq.~(\ref{eq:S3.10}) are: $k_0 = 4.05~\mu$m$^{-1}$, $n'_{\mathrm{eff}} = 1.86\cdot 10^{-4}$~K$^{-1}$, $L=300~\mu$m, $R_L=1.47$~k$\Omega$, $G=5\cdot 10^{6}$~W$\cdot$K$^{-1}\cdot$m$^{-2}$, and $A=2.25\cdot 10^{-8}$~m$^2$. The estimated distribution of $\Delta\phi_e$ exhibits a variance of $3\cdot 10^{-11}$. Measurements performed for other values of $I_0$ lead to a variance of $\Delta\phi_e$ of the same order of magnitude, thus allowing us to consider this phase noise negligible.

In large-scale programmable integrated photonic (PIP) circuits, composed of meshes of Mach-Zehnder interferometers [7], the global phase noise induced by thermo-optic phase shifters is intrinsically bounded. Let us assume that each phase shifter introduces an independent random phase error $\Delta\phi_e$ with variance $\sigma_1^2 \sim 10^{-11}$. In the worst-case scenario, corresponding to $N$ phase shifters connected in series, the accumulated phase error is given by the sum of $N$ independent and identically distributed random variables, which leads to a total variance $\sigma_N^2 = N\sigma_1^2$ [9]. However, in realistic PIP meshes the phase shifters are embedded within interferometric structures, and their contributions to the relative phase between two output ports are weighted by the corresponding transfer matrices of the interferometers. Consequently, the effective phase error affecting such relative phase necessarily exhibits a variance smaller than or equal to that of the worst-case serial configuration. Therefore, the phase noise that perturbs the differential phase of an anbit at the output of a PIP mesh integrating a set of $N$ phase shifters is upper-bounded by the variance obtained when all phase errors accumulate sequentially along the same optical path, i.e., $\sigma_N^2 \leq N\sigma_1^2$. This upper bound reaches values of the same order of magnitude as the relative intensity noise (RIN) contributions in the elevation and azimuthal angles of the GBS ($\sim 10^{-5}$) only for $N \gtrsim 10^6$, which exceeds the number of phase shifters that can currently be integrated in a PIP platform [10]. Consequently, \emph{the phase noise induced by thermo-optic phase shifters can be safely considered negligible}.

\section{RIN, shot, and thermal noise models}

The combined contributions of RIN, shot noise, and thermal noise in $\mathrm{Var}(\delta I_k)$ (with $k\in\{0,1,\varphi\}$) are calculated as the sum of three independent noise sources:
\begin{equation}
\mathrm{Var}(\delta I_k) = \mathrm{Var}(\delta I_k)_{\mathrm{RIN}} + \mathrm{Var}(\delta I_k)_{\mathrm{shot}} + \mathrm{Var}(\delta I_k)_{\mathrm{thermal}},
\label{eq:S4.1}
\end{equation}
where the individual noise contributions are calculated by using the well-known semi-classical models reported in the optical literature [11]:
\begin{equation}
\mathrm{Var}(\delta I_k)_{\mathrm{RIN}} = \mathrm{RIN}\cdot B \cdot \overline{I_k}^2,
\label{eq:S4.2}
\end{equation}
\begin{equation}
\mathrm{Var}(\delta I_k)_{\mathrm{shot}} = 2\cdot q \cdot B \cdot \overline{I_k},
\label{eq:S4.3}
\end{equation}
\begin{equation}
\mathrm{Var}(\delta I_k)_{\mathrm{thermal}} = \frac{4\cdot k_B \cdot T \cdot B}{R_L},
\label{eq:S4.4}
\end{equation}
where RIN is expressed in 1/Hz (linear units), $B$ is the bandwidth of the photodiodes (20 GHz), $q$ is the elementary charge, $k_B$ is the Boltzmann constant, $T$ is the temperature, and $R_L$ is the load resistance (680 $\Omega$). In addition, note that $\overline{I_0}$, $\overline{I_1}$, and $\overline{I_\varphi}$ (the ideal photocurrents) can be expressed as a function of $\overline{r}$, $\overline{\theta}$ and $\overline{\varphi}$ (the ideal values of the EDFs) by using Eqs.~(3) and (5) of the paper:
\begin{equation}
\overline{I_0} = \frac{\mathcal{R}}{2}\overline{r}^2\cos^2\frac{\overline{\theta}}{2},
\label{eq:S4.5}
\end{equation}
\begin{equation}
\overline{I_1} = \frac{\mathcal{R}}{2}\overline{r}^2\sin^2\frac{\overline{\theta}}{2},
\label{eq:S4.6}
\end{equation}
\begin{equation}
\overline{I_\varphi}_{\mathrm{Unbalanced}} = \frac{\mathcal{R}}{4}\overline{r}^2\left(1-\sin\overline{\theta}\sin\overline{\varphi}\right),
\label{eq:S4.7}
\end{equation}
\begin{equation}
\overline{I_\varphi}_{\mathrm{Balanced}} = -\frac{\mathcal{R}}{2}\overline{r}^2\sin\overline{\theta}\sin\overline{\varphi}.
\label{eq:S4.8}
\end{equation}

Finally, combining Eqs.~(\ref{eq:S4.1})-(\ref{eq:S4.4}) with Eqs.~(\ref{eq:S4.5})-(\ref{eq:S4.8}) and substituting them into Eqs.~(6)-(9) of the paper, we obtain after some algebraic manipulation the following closed-form expressions:
\begin{equation}
\mathrm{Var}(r) = \frac{1}{\mathcal{R}^2\overline{r}^2}\left[\frac{8 k_B T B}{R_L} + \mathcal{R}\overline{r}^2 q B + \frac{\mathcal{R}^2 B\, \mathrm{RIN}\, \overline{r}^4}{4}\left(1-\tfrac{1}{2}\sin^2\overline{\theta}\right)\right],
\label{eq:S4.9}
\end{equation}
\begin{equation}
\mathrm{Var}(\theta) = \frac{B\,\mathrm{RIN}}{2}\sin^2\overline{\theta} + \frac{4qB}{\mathcal{R}^2\overline{r}^2} + \frac{8 k_B T B}{R_L \mathcal{R}^2 \overline{r}^4}\left[\tan^2\frac{\overline{\theta}}{2}+\cot^2\frac{\overline{\theta}}{2}\right],
\label{eq:S4.10}
\end{equation}
\begin{multline}
\mathrm{Var}(\varphi)_{\mathrm{Unbalanced}} = \frac{1}{\sin^2\overline{\theta}\cos^2\overline{\varphi}}\Bigg\{B\,\mathrm{RIN}\left[2-\tfrac{1}{2}\sin^2\overline{\theta}-3\sin\overline{\theta}\sin\overline{\varphi}+\tfrac{3}{2}\sin^2\overline{\theta}\sin^2\overline{\varphi}\right] \\
+ \frac{4qB}{\mathcal{R}\overline{r}^2}\left[3-4\sin\overline{\theta}\sin\overline{\varphi}+\sin^2\overline{\varphi}\right] \\
+ \frac{16 k_B T B}{R_L \mathcal{R}^2 \overline{r}^4}\left[6 - \frac{4\sin\overline{\varphi}}{\sin\overline{\theta}}+2\sin^2\overline{\varphi}\left(\frac{2}{\sin^2\overline{\theta}}-1\right)\right]\Bigg\},
\label{eq:S4.11}
\end{multline}
\begin{multline}
\mathrm{Var}(\varphi)_{\mathrm{Balanced}} = \frac{1}{\sin^2\overline{\theta}\cos^2\overline{\varphi}}\Bigg\{\frac{B\,\mathrm{RIN}}{2}\left[1+2\sin^2\overline{\theta}\sin^2\overline{\varphi}\right] \\
+ \frac{4qB}{\mathcal{R}\overline{r}^2}\left[1+\sin^2\overline{\varphi}\right] \\
+ \frac{16 k_B T B}{R_L \mathcal{R}^2 \overline{r}^4}\left[2+\sin^2\overline{\varphi}\left(\frac{4}{\sin^2\overline{\theta}}-2\right)\right]\Bigg\}.
\label{eq:S4.12}
\end{multline}

Remarkably, Eqs.~(\ref{eq:S4.9})-(\ref{eq:S4.12}) provide the complete information about the variance of each EDF across the GBS as a function of the mean values of the EDFs. These expressions are used to calculate Fig.~3 of the main text and predict some interesting features in the behavior of the EDFs. For example, the harmonic dependence of $\mathrm{Var}(r)$ in the presence of RIN can be directly inferred from Eq.~(\ref{eq:S4.9}) by neglecting the shot and thermal noise contributions.

\section{Manufacturing, assembly, and characterization of the PIP circuit}

\noindent\textbf{Manufacturing process}

The PIP used for experimental validation was fabricated by Advanced Micro Foundry on a silicon-on-insulator platform. The process employs a 220 nm thick silicon device layer, patterned using 193 nm deep ultraviolet lithography to define single-mode waveguides with a nominal width of 500 nm.

Thermo-optic phase shifters are implemented by integrating metallic micro-heaters directly above the waveguides. These heaters consist of a 120 nm titanium nitride layer deposited over the silicon core. In order to enhance thermal efficiency and reduce power consumption, the phase-shifter sections incorporate suspended waveguide geometries and etched isolation trenches. This configuration enables a measured tuning efficiency of approximately 1.35 mW per $\pi$ phase shift.

Photodetection is realized through germanium-on-silicon photodiodes, featuring 20 GHz bandwidth, integrated within the same platform. The devices exhibit responsivities up to 0.85 A/W, allowing accurate recovery of the anbit states at the opto-electrical conversion stage.\\

\noindent\textbf{Assembly process}

Post-fabrication, the PIC was mounted onto a custom-designed printed circuit board and electrically packaged via wire bonding. This arrangement allows independent control of each thermo-optic phase shifter and individual access to the integrated photodiodes.

Optical excitation is provided through vertical fiber-to-chip coupling using surface grating couplers. No permanent optical packaging was implemented. The grating couplers are optimized around a wavelength of 1550 nm and exhibit insertion losses of approximately 4 dB per grating.

Due to the short optical path lengths inside the chip, internal propagation losses are negligible for the present experiments. For reference, the typical waveguide propagation loss of the platform is approximately 1.17 dB/cm, which does not significantly impact the measured constellations.\\

\noindent\textbf{Characterization}

Each thermo-optic phase shifter was independently calibrated by monitoring the optical response with the integrated photodetectors. The extracted phase shift follows the expected quadratic dependence on the applied current, consistent with the Joule-heating mechanism governing thermo-optic modulation.

Optical power transfer was fitted using the standard input--output relations of a Mach--Zehnder interferometer operated in either bar or cross configuration, enabling accurate determination of extinction ratios and phase tuning curves.

During system-level characterization, the phase shifters were driven by programmable multichannel current sources. Photocurrent acquisition was performed using high-precision source measurement units configured with NPLC = 1 to ensure stable statistical sampling.

\section{Total angular variance}

The metric line element in spherical coordinates is given by the expression [12]:
\begin{equation}
\mathrm{d}s^2\rfloor_{\mathrm{3D}} = \mathrm{d}r^2 + r^2\mathrm{d}\theta^2 + r^2\sin^2\theta\,\mathrm{d}\varphi^2.
\label{eq:S6.1}
\end{equation}
By restricting $\mathrm{d}s^2$ to the angular dimensions around the ideal state with coordinates $(\overline{r},\overline{\theta},\overline{\varphi})$, we obtain that:
\begin{equation}
\mathrm{d}s^2\rfloor_{\mathrm{2D}} = \overline{r}^2\mathrm{d}\theta^2 + \overline{r}^2\sin^2\overline{\theta}\,\mathrm{d}\varphi^2.
\label{eq:S6.2}
\end{equation}
Since the first-order EPT establishes that $\mathrm{d}\theta \approx \delta\theta$ and $\mathrm{d}\varphi \approx \delta\varphi$ (see Supplementary Section 2), noise-induced errors in both the elevation and azimuthal angles can be modeled by the random variable:
\begin{equation}
(\delta s)^2 \coloneqq \overline{r}^2(\delta\theta)^2 + \overline{r}^2\sin^2\overline{\theta}\,(\delta\varphi)^2.
\label{eq:S6.3}
\end{equation}
Accordingly, the total angular variance ($\sigma_{\mathrm{T}}^2$) can be defined as the \emph{mean-square angular deviation}, that is, the expectation value of $(\delta s)^2$:
\begin{align}
\sigma_{\mathrm{T}}^2 &= \mathbb{E}\left[(\delta s)^2\right] = \overline{r}^2\,\mathbb{E}\left[(\delta\theta)^2\right] + \overline{r}^2\sin^2\overline{\theta}\,\mathbb{E}\left[(\delta\varphi)^2\right] \nonumber \\
&\equiv \overline{r}^2\mathrm{Var}(\theta) + \overline{r}^2\sin^2\overline{\theta}\,\mathrm{Var}(\varphi),
\label{eq:S6.4}
\end{align}
which matches the closed-form expression used in the main text.

\section*{Supplementary references}

\begin{enumerate}[label={[\arabic*]},leftmargin=*,itemsep=2pt]

\item A. Macho-Ortiz, R. López-March, P. Martínez-Carrasco, F. J. Fraile-Peláez, and J. Capmany, ``Analog programmable-photonic information,'' Advanced Photonics \textbf{8}, 036002 (2026).

\item R. Syms and J. Cozens, \emph{Optical Guided Waves and Devices}. (McGraw-Hill, London, 1992).

\item L. B. Soldano and E. C. M. Pennings, ``Optical multi-mode interference devices based on self-imaging: principles and applications,'' Journal of Lightwave Technology \textbf{13}, 615 (1995).

\item F. A. Morrison, \emph{Uncertainty Analysis for Engineers and Scientists: A Practical Guide}. (Cambridge University Press, Cambridge, 2021).

\item K. Petermann, \emph{Laser Diode Modulation and Noise}. (Kluwer Academic Publishers, London, 1988).

\item A. Macho-Ortiz, D. Pérez-López, J. Azaña, and J. Capmany, ``Analog programmable-photonic computation,'' Laser \& Photonics Reviews \textbf{17}, 2200360 (2023).

\item J. Capmany and D. Pérez, \emph{Programmable Integrated Photonics}. (Oxford University Press, Oxford, 2020).

\item M. Jacques, A. Samani, E. E.-Fiky, D. Patel, Z. Xing, and D. V. Plant, ``Optimization of thermo-optic phase-shifter design and mitigation of thermal crosstalk on the SOI platform,'' Optics Express \textbf{27}, 10456 (2019).

\item H. P. Hsu, \emph{Probability, Random Variables, \& Random Processes}. (McGraw-Hill, 1997).

\item D. Pérez-López and L. Torrijos-Morán, ``Large-scale photonic processors and their applications,'' npj Nanophotonics \textbf{2}, 32 (2025).

\item G. P. Agrawal, \emph{Fiber-Optic Communication Systems}. (Wiley, New Jersey, 2010).

\item M. Fecko, \emph{Differential Geometry and Lie Groups for Physicists}. (Cambridge University Press, Cambridge, 2006).

\end{enumerate}

\end{document}